\documentclass{jfm}
\usepackage{graphicx}
\usepackage{epstopdf,epsfig}
\usepackage{amsmath}
\usepackage{natbib}
\usepackage{hyperref}
\usepackage{subcaption}
\usepackage{IEEEtrantools}
\hypersetup{
    colorlinks = true,
    urlcolor   = red,
    citecolor  = blue,
}
\graphicspath{{Figs2Sync/},{Figs/}}

\newcommand{\e}{\mathrm{e}}
\newcommand{\ud}{\;\mathrm{d}}

\newcommand{\add}[1]{{\color{black}#1}}

\renewcommand{\i}{\mathrm{i}}
\newcommand{\RomanNumeralCaps}[1]

\title{Low-wavenumber wall pressure fluctuations in turbulent flows within concentric annular ducts}

\author{Yaomin Zhao\aff{1,2}, Taiyang Wang\aff{1,2}
\and Benshuai Lyu\aff{1}
 \corresp{\email{b.lyu@pku.edu.cn}}
 }
\affiliation{\aff{1}State Key Laboratory of Turbulence and Complex Systems, College of Engineering, Peking University, Beijing 100871, China.
\aff{2}HEDPS, Center for Applied Physics and Technology, College of Engineering, Peking University, Beijing 100871, China.
}

\begin{document}
\maketitle
\begin{abstract}
Compressible direct numerical simulations of turbulent channel flows in
concentric annular ducts of height $2\delta$ are performed to study the
low-wavenumber wall pressure fluctuations (WPF) over cylindrical walls \add{at a
bulk Mach number $M_b = 0.4$ and bulk Reynolds number $Re_b=3000$.} The
radius of the inner cylinder $R$ is varied between $0.2\delta$, $\delta$,
$2\delta$ and $\infty$. As $R$ decreases, the one-point power spectral density
of the WPF decreases at intermediate but increases at high frequencies. When $R$
decreases, the 1D (streamwise) wavenumber-frequency spectrum of the WPF
decreases at high wavenumbers. At low wavenumbers, however, as $R$ reduces to
$0.2\delta$ the 1D wavenumber-frequency spectrum exhibits multiple spectral
peaks whose strengths increase with frequency. Examination of the 2D
wavenumber-frequency spectra \add{shows that these represent \add{acoustic duct}
modes} that closely match theoretical predictions. \add{The acoustic modes of
higher radial orders exhibit increasingly high amplitude on the inner than on the
outer walls.} The low-wavenumber components of the $0$th-order (azimuthal) 2D
wavenumber-frequency spectrum are of great importance in practice, and their
magnitude increases as $R$ reduces; this increase is increasingly pronounced at
higher frequencies. Analytical modelling \add{and numerical validation show}
that this increase appears to arise from the \add{``geometric'' effects
connected with the Green's function}, and they are generated mainly by radial
and azimuthal disturbances. Disturbances closer to the wall are shown to be
increasingly important in WPF generation as $R$ reduces, which highlights a
potential in WPF control using wall treatment on thin cylinders.
\end{abstract}


\begin{keywords}
\end{keywords}

\section{Introduction}
\label{sec:introduction}
Turbulent flows exert unsteady wall pressure loadings over solid walls. The wall pressure fluctuations (WPF) play an essential role in \add{sonic loading in rocket sensor fairings, flow noise inside towed sonar arrays, and flow-induced vibration of pipelines}. A clear characterisation and understanding of their statistics, spectra, and control are crucial in the aviation and ocean engineering industry~\citep{blake_turbulent_1970}.

Extensive research has been conducted to characterise the WPF beneath turbulent boundary layers, with early research mainly on incompressible flows over flat plates. \citet{Willmarth1956} appeared to be the first to measure basic statistics such as the root mean square (rms) value and frequency spectra of the WPF. Further experiments conducted by \citet{Harrison1958} and \citet{Willmarth1958} studied the WPF's space-time correlation and showed that the convection velocity was around $80\%$ of the freestream velocity. \citet{Bull1967}, on the other hand, reported that the convection velocity depends on frequency.
Regarding analytical models, \citet{kraichnan_pressure_1956} was among the first to develop a mirror-flow model and showed that the rms value of the WPF was roughly six times the mean wall shear stress, a result that agreed well with the measured values by \citet{Willmarth1956} and many others~\citep{Bull1996}. Based on streamwise and lateral correlation similarity, \citet{Corcos1963} proposed a model for the cross-spectral density of the WPF, applicable to the wavenumber regime near $\omega/U_c$, where $\omega$ and $U_c$ represent the angular frequency and convection velocity of the WPF, respectively. Comprehensive reviews of these early studies can be found in \citet{Willmarth1975} and \citet{Bull1996}. 



However, realistic applications often involve wall surfaces with curvature, such as the hose of a towed sonar array. Incompressible WPF over curved surfaces was studied subsequently. This includes, for example, early experiments by \citet{Willmarth1970} and \citet{Willmarth1976}. They concluded that the wall curvature caused insignificant changes in the frequency spectra and convection velocity but a noticeable reduction in the spanwise correlation length. Analytical modelling was conducted by \citet{Howe1987} and \citet{dowling_underwater_1998}, the latter of which modelled the incompressible wavenumber-frequency spectrum using Lighthill's acoustic analogy. The effect of wall curvature on WPF was also examined using direct numerical simulations (DNS), where \citet{Neves1994b} showed that increasing curvature reduced both spatial and temporal spectra of the WPF at all scales, decreased the azimuthal correlation length, increased the streamwise correlation length, and reduced the convection velocity. Note that the curvature examined in their studies was substantially larger than that in \citet{Willmarth1970} and \citet{Willmarth1976}. 

\begin{figure}
    \centering
    \includegraphics[width=0.8\textwidth]{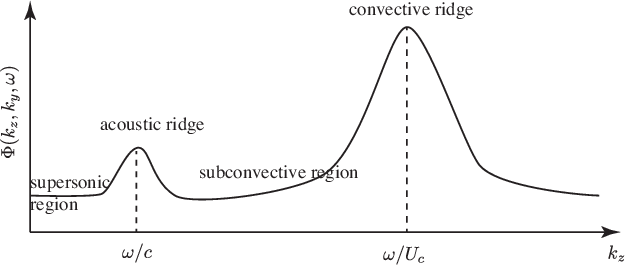}
    \caption{\add{A typical wavenumber-frequency spectrum of the WPF at a given
    frequency $\omega$ and spanwise wavenumber $k_y=0$. $k_z$ represents the streamwise wavenumber}.}
    \label{fig:typicalSpectrum}
\end{figure}

A typical wavenumber-frequency spectrum at a given frequency $\omega$ is shown in figure~\ref{fig:typicalSpectrum}. A high-wavenumber peak called convective ridge is located at $k_z = \omega/ U_c$ due to turbulence convection, \add{where $k_z$ represents the streamwise wavenumber.} In contrast, a low-wavenumber acoustic ridge appears at $k_z = \omega / c_0$, where $c_0$ is the speed of sound. Note that, however, it is the low-wavenumber components of the WPF, i.e., the subconvective and supersonic regions shown in figure~\ref{fig:typicalSpectrum}, that are of particular interest in practice. This is because they are more likely to couple with the elastic motion of the solid walls or are responsible for direct acoustic radiation. In such regimes, compressibility is crucial, even in very low Mach number flows; it is therefore essential to include the effects of compressibility in the study of WPF. For example, using Lighthill's acoustic analogy, \citet{williams_surface-pressure_1965} showed that including compressibility resulted in the violation of the Kraichnan-Phillips theorem and led to a non-zero spectral magnitude proportional to the square of the mean-flow Mach number at zero wavenumbers. 




The magnitude of the wavenumber-frequency spectrum around the acoustic ridge, as can be seen from figure~\ref{fig:typicalSpectrum}, is often much lower than that near the convective ridge. This poses significant challenges to resolving the low-wavenumber components using either experiments or numerical simulations. As such, compressible simulations that study the subconvective and supersonic regions remain scarce. \citet{Gloerfelt2013} and \citet{Cohen2018} were among the first to study the acoustic component of the WPF beneath a turbulent boundary layer of Mach 0.5 using large eddy simulations (LES). They were able to capture the acoustic ridge, but were unable to fully eliminate errors caused by artificial boundaries. A recent study by \citet{Liu2024} showed that with suitable choices of low-dissipative and low-dispersive schemes, it is possible to accurately resolve these low-energy components using compressible DNS. The DNS was performed on a rectangular channel with periodic boundary conditions to suppress the artefacts introduced by the boundary conditions. 

However, as mentioned above, practical applications are rarely characterised by flat plates, but rather by surfaces with curvature. To the best of the authors' knowledge, compressible flow simulations focusing on the low-wavenumber region of the WPF over surfaces with curvature remain to be seen. The characteristics of the low-wavenumber components of the WPF over cylindrical walls, in particular, how energetic they are compared to convective components as the curvature varies, remain largely unknown. In this paper, we aim to bridge the gap and study the wall pressure over cylindrical surfaces of various curvatures. To directly resolve these low-frequency components, we resort to DNS with low-dissipative and low-dispersive finite-difference schemes. Similar to \citet{Liu2024}, to avoid the inevitable noise introduced by artificial boundary conditions, channel flows within a concentric annular duct are simulated with a periodic boundary condition in the streamwise direction. In doing so, not all the features of the free boundary layer can be captured, but the important effects of curvature are kept without complications introduced by simulating free boundary layers~\citep{Neves1994b}. 

This paper is structured as follows. \add{Section 2 introduces the computational
setup and discretisation schemes used in this study. Section 3 validates the
simulations and discusses the effects of curvature on flow statistics. Section 4
shows the power spectral density (PSD) and wavenumber-frequency spectra of the
WPF over cylinders of various diameters. Section 5 uses Lighthill's acoustic
analogy to model the WPF generated by cylindrical boundary layers, with which
the mechanisms of the low-wavenumber WPF augmentation are discussed.} The
following section concludes the paper and lists future work.

\section{Numerical setup}
\label{sec:setup}

\begin{table}
  \begin{center}
\def~{\hphantom{0}}
  \begin{tabular}{ccccccccc}
      Case & $Re_b$ & \begin{tabular}{@{}c@{}}$Re_{\tau}$\\$(\mathrm{inner},
      \mathrm{outer})$\end{tabular} &$\epsilon$ & $L_{\theta},L_r,L_z$ &
      $N_{\theta}, N_r, N_z$ & $\Delta_\theta^+$  & $\Delta_r^+$  & $\Delta_z^+$ \\
      1 & 3000& $188$& $\infty$ & $\frac{4}{3}\pi \delta , 2\delta ,16\pi \delta$ & $ 256 , 193 , 1024
      $ & $3.08$ & $0.20-2.40$ & $9.23$ \\
      2 & 3000& $193,183$& $2$ & $4\pi \delta , 2\delta , 16\pi \delta$ & $800 , 193 , 1024$ & $3.11-6.23$ & $0.43-5.8$ & $9.73$ \\
      3 & 3000& $198,183$& $1$ & $ 2\pi \delta , 2\delta , 16\pi \delta$ & $400 , 193 , 1024$ & $3.20-9.59$ & $0.25-6.71$ & $9.99$ \\
      4 & 3000& $230,182$& $0.2$ & $\frac{2}{5}\pi \delta , 2\delta , 16\pi \delta$ & $300 , 193 , 1024$ & $0.99-10.87$ & $0.56-6.71$ & $11.58$ 
  \end{tabular}
  \caption{\add{Simulation matrix and parameters used in each simulation. $N_i$
  and $\Delta^+_i$ represent the number and non-dimensional of the grid element
  in $i$ direction ($i$ = $\theta$, $r$ and $z$).}}
 \label{tab:simulationMatrix}
  \end{center}
\end{table}

 \begin{figure}
     \centering
     \includegraphics[width=0.6\linewidth]{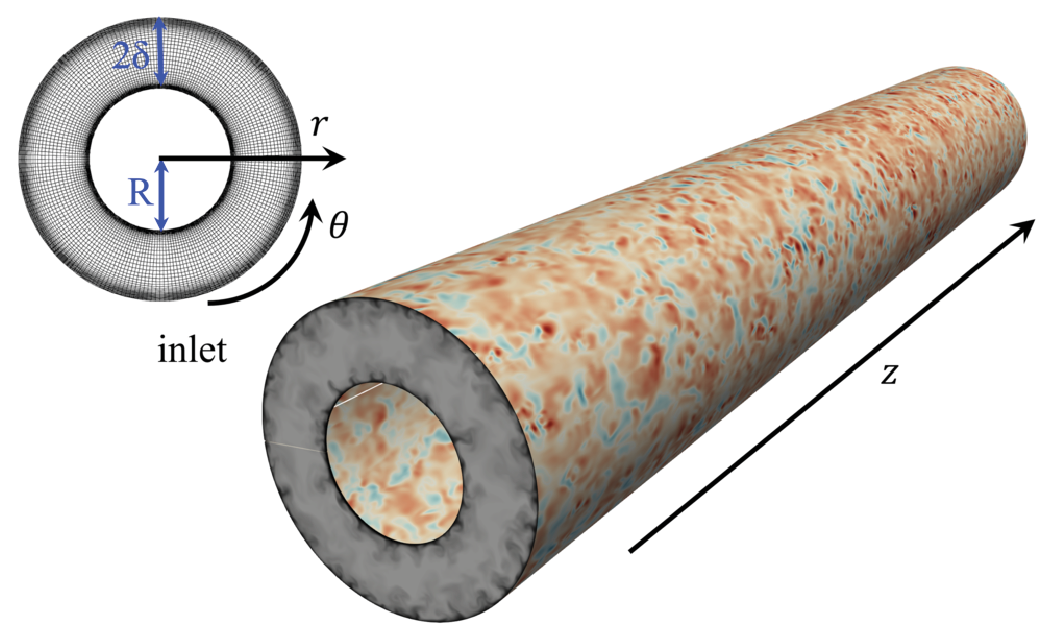}
    \caption{\add{DNS configurations, with $\theta$, $r$, and $z$ denoting the
    circumferential, radial, and axial directions, respectively. The contours of
    instantaneous velocity magnitude and pressure are shown on the inlet and
    inner and outer walls, respectively.}}
    \label{fig:case_config}
\end{figure}

A schematic setup of the simulations is shown in figure~\ref{fig:case_config}.
The channel flow within an annular duct is \add{discretised} on a generalised
curvilinear coordinate system, with $\theta$, $r$ and $z$ denoting the
azimuthal, radial, and axial directions, respectively. $L_r$ and $L_z$ represent
the channel height and length, while $R$ and $L_\theta$ denote the radius and
the circumference of the inner cylindrical wall, respectively. In the $\theta$
and $z$ directions, grid points are uniformly distributed, with periodic
conditions applied at the boundaries. A no-slip isothermal condition is imposed
on the walls, and the grid near the wall is refined based on a hyperbolic
tangent function. The parameters of all four DNS simulations are listed in
Tab.~\ref{tab:simulationMatrix}, where $\epsilon$ is defined as $R/\delta$ to
represent the non-dimensional radius of the inner cylinder. Case 1 is a plane
channel for comparison, where $L_\theta$ reduces to the spanwise domain size.
\add{Note that the channel length $L_z$ is deliberately chosen to be
$16\pi\delta$ in order to resolve the low-wavenumber components of the WPF.}

\add{The code HiPSTAR \citep{sandberg2015compressible}, which has been extensively
validated for acoustic simulations \citep{deuse2020different}, is used to solve
the non-dimensionalized three-dimensional (3D) compressible Navier-Stokes
equations

\begin{align}
    \frac{\partial \rho}{\partial t} + \frac{\partial (\rho u_j)}{\partial x_j} &= 0, \tag{2.1a} \\
    \frac{\partial (\rho u_i)}{\partial t} + \frac{\partial (\rho u_i u_j + p\delta_{ij})}{\partial x_j} &= \frac{\partial \tau_{ij}}{\partial x_j}, \tag{2.1b} \\
    \frac{\partial (\rho e_0)}{\partial t} + \frac{\partial [u_j(\rho e_0 + p)]}{\partial x_j} &= \frac{\partial (\tau_{ij} u_i)}{\partial x_j} - \frac{\partial q_j}{\partial x_j}. \tag{2.1c}
\end{align}
Here, \(\rho\), \(u_i\), \(p\) and \(T\) denote the non-dimensionalised flow density, velocity components, pressure and temperature, respectively. 

The non-dimensionalization results in dimensionless parameters as \(Re_b = (\rho_b U_b \delta)/\mu_\infty = 3000\) and \(M_b = U_b / c_\infty = 0.4\), where \(\delta\), \(U_b\) and \(\rho_b\)  are the channel half-height, mean bulk velocity and density, respectively. Moreover, \(\mu_\infty\) and \(c_\infty\) are viscosity and acoustic velocity for the reference state which are only dependent on the reference wall temperature \(T_w\). The total energy \(e_0\) is given by 
\begin{equation}
    e_0 = \frac{\rho u_i u_i}{2} + \frac{1}{\gamma (\gamma - 1) M_b^2}T,
\end{equation}
where \(\gamma = 1.4\) is the specific heat ratio. Moreover, the stress tensor is written as 
\begin{equation}
    \tau_{ij} = \frac{\mu}{Re_b}\left(\frac{\partial u_i}{\partial x_j} + \frac{\partial
    u_j}{\partial x_i} - \frac{2}{3}\frac{\partial u_k}{\partial
    x_k}\delta_{ij}\right),
\end{equation}
and the molecular viscosity \(\mu\) is computed using Sutherland's law
(\citep{white1991viscous}. Similarly, the heat flux \(q_j\) is written as  \begin{equation}
    q_j = -\frac{\mu}{(\gamma-1) Pr Re_b M_b^2}\frac{\partial T}{\partial x_j},
\end{equation}
where $Pr$ is the Prandtl number and takes the value of $0.72$. }

\add{A fourth-order finite difference scheme was applied for spatial
discretisation, while the ultra-low storage frequency optimised explicit
Runge-Kutta method \citep{kennedy2000low} was used for time integration. In
addition, the non-linear terms are processed with a skew-symmetric splitting
method \citep{kennedy2008reduced} to reduce aliasing, and a standard filtering
scheme \citep{bogey2004family} is used to improve numerical stability.
Simulations are conducted for around 48 flow-through time units on GPUs, after
which the wall pressure on the inner cylinder surface and velocity data
within the channel are stored for analysis. Detailed sampling frequency and
sampling duration are shown in section~\ref{sec:Results}.}


\section{Validation and flow statistics}
\label{sec:validationStatistics}
\add{Before examining the WPF, we first validate the present DNS by
comparing the flow statistics against those in the literature. The effects of
curvature on the flow statistics are shown subsequently. To facilitate an
examination of the turbulent statistics distribution along the wall-normal direction, a radial distance measured from the inner cylinder wall, defined as
$y=r - R$, is used. This distance is often presented in wall units defined by
$y^+=yu_\tau/\nu$, where $\nu$ and $u_\tau$ represent the kinematic viscosity
and wall friction velocity of the flow, respectively.

\label{sec:validation}
\subsection{Validation of the DNS}
To validate the present simulations, the mean and fluctuation velocities are
first compared against results published previously. Note that the fluctuation
velocities are defined by removing the mean values defined by the Favre
average ($\tilde{\cdot}$), i.e.
\begin{IEEEeqnarray}{rCl}
    \tilde{u}_i &=& \langle \rho u_i \rangle / \langle \rho \rangle,
    \IEEEyessubnumber \\
    u_i^\prime &=& u_i - \tilde{u}_i, \IEEEyessubnumber
\end{IEEEeqnarray}
where $\langle\cdot\rangle$ represents ensemble average. In practice, temporal
and spatial (streamwise and azimuthal) averages are used instead. It is worth
noting that although the bulk Mach number $M_b$ is $0.4$, the density variation of the flow is very limited. As such, the difference between the direct and Favre averages is negligible. The mean and fluctuation velocities are
subsequently normalised by the friction velocity $u_\tau$ to obtain
$\tilde{u}_i^+$ and $u_i^{\prime+}$, respectively.} 

\begin{figure}
    \centering
    \begin{subfigure}{0.495\textwidth}
    \includegraphics[width=\linewidth]{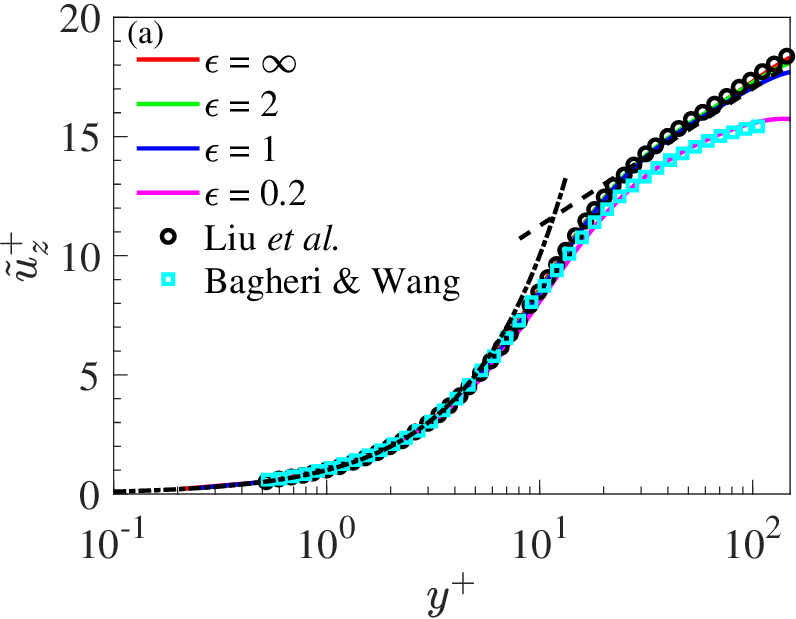}
    \end{subfigure}
    \begin{subfigure}{0.495\textwidth}
    \includegraphics[width=\linewidth]{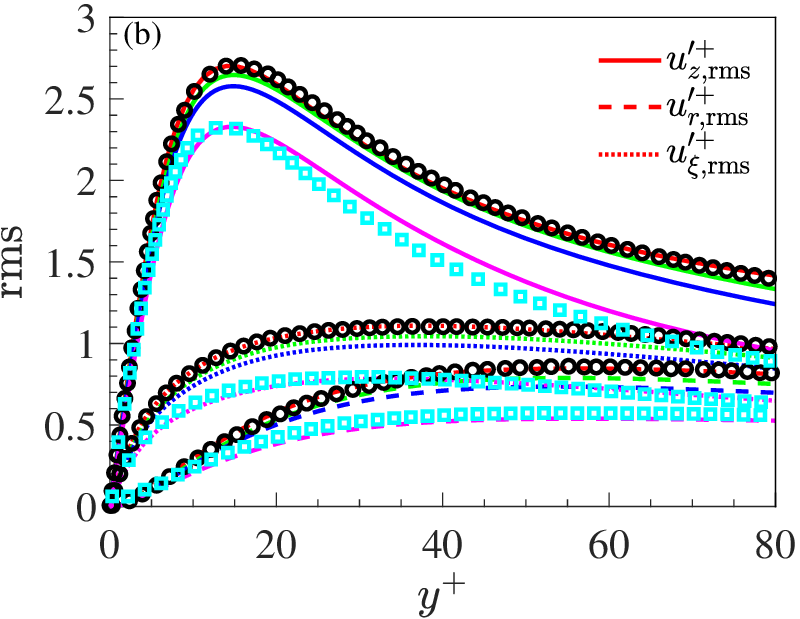}
    \end{subfigure}
    \caption{Comparison of (a) mean velocity profiles and (b) rms of velocity
    fluctuations between the present cases and previous simulations. Circle,
    plane channel case in \citet{Liu2024}; square, case with $\epsilon=0.22$ in
    \citet{BAGHERI2020}.}
    \label{fig:val}
\end{figure}

\add{The comparison of the two velocities is shown in figure~\ref{fig:val},
among which figure~\ref{fig:val}(a) shows} the distribution of the axial mean
velocity along the wall-normal direction. \add{In the rectangular case, the viscous
sublayer and log law of the wall are captured and compared well to
\citet{Liu2024}.} When the radius of the inner cylinder reduces to $\epsilon=2$
or $1$, the velocity distribution shows minimal change, in contrast to the
significant changes at $\epsilon=0.2$. In particular, the axial mean velocity
decreases in the outer layer. The obtained velocity profile agrees well with
\citet{BAGHERI2020}, albeit the slight mismatches in the inner cylinder radius.

Figure~\ref{fig:val}(b) shows the rms value of the velocity fluctuations in the
axial, radial, and azimuthal directions, respectively. In the rectangular case,
we see that the velocity fluctuations compare very well with \citet{Liu2024}.
When the inner radius decreases, the velocity fluctuations continuously
decrease, particularly in the outer layer of the flow. \add{Note that the study
of \citet{BAGHERI2020} was conducted for a $\epsilon=0.22$, hence it is
different from the present case of $\epsilon=0.2$. This partly explains why
there are slight mismatches in the rms values of the two cases. In addition, the
difference in the Mach number might also play a role as incompressible
simulations were conducted in \citet{BAGHERI2020}. Nevertheless,
figure~\ref{fig:val}(b) still shows a relatively good agreement between the two
cases.} In light of the good agreement shown in figure~\ref{fig:val}, one
expects that the current simulation captures the essential physics of turbulent
channel flows within concentric \add{annular} ducts.

\add{\subsection{Effects of curvature on the flow field} 
To examine the effects of curvature on the flow closely, we show the distribution of the Reynolds stresses along the wall normal direction for various cylinder diameters. These are shown in figure~\ref{fig:ReynoldsInner}. Figure~\ref{fig:ReynoldsInner}(a) shows the wall-normal distribution of $\langle u^{\prime+}_zu^{\prime+}_z\rangle$. One sees that this Reynolds stress component peaks at around $y^+=15$, while increasing the curvature appears to move this location slightly closer to the wall. This finding is consistent with the incompressible simulation by \citet{Neves1994a}. As the curvature increases from $\epsilon=\infty$ to $\epsilon=1$, the amplitude of this component decreases slightly. As the curvature increases to $\epsilon=0.2$ significant reduction occurs. This suggests that axial velocity fluctuations tend to be suppressed within boundary layers over thinner cylinders.
\begin{figure}
    \centering
    \begin{subfigure}{0.495\textwidth}
    \includegraphics[width=\linewidth]{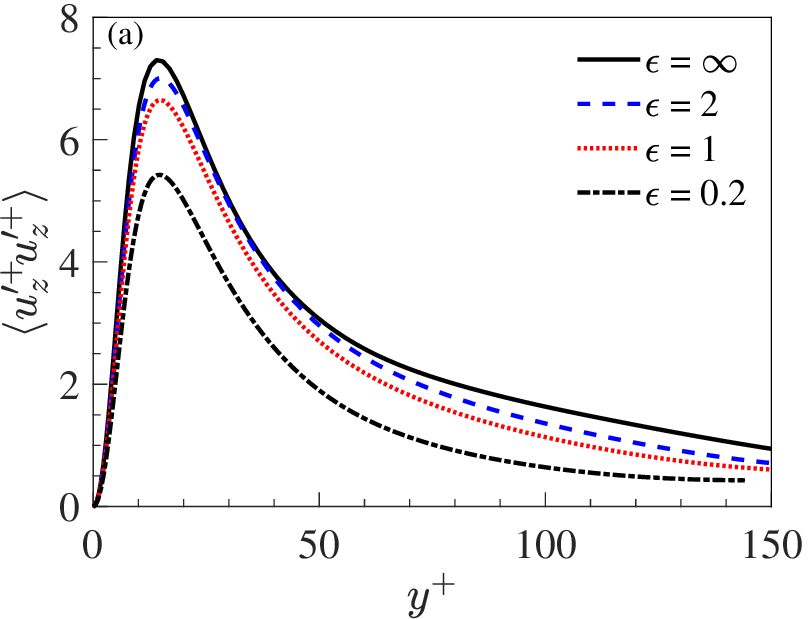}
    \end{subfigure}
    \begin{subfigure}{0.495\textwidth}
    \includegraphics[width=\linewidth]{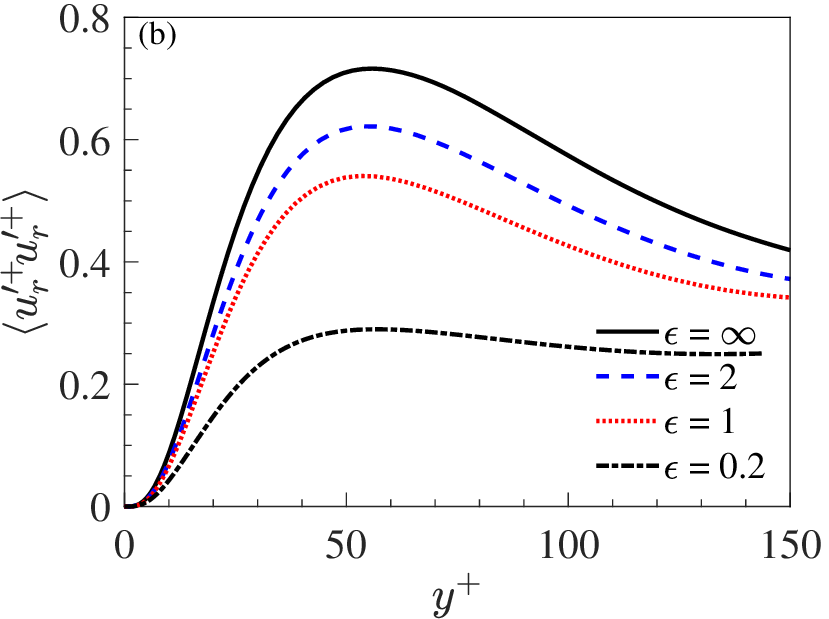}
    \end{subfigure}
    \begin{subfigure}{0.495\textwidth}
    \includegraphics[width=\linewidth]{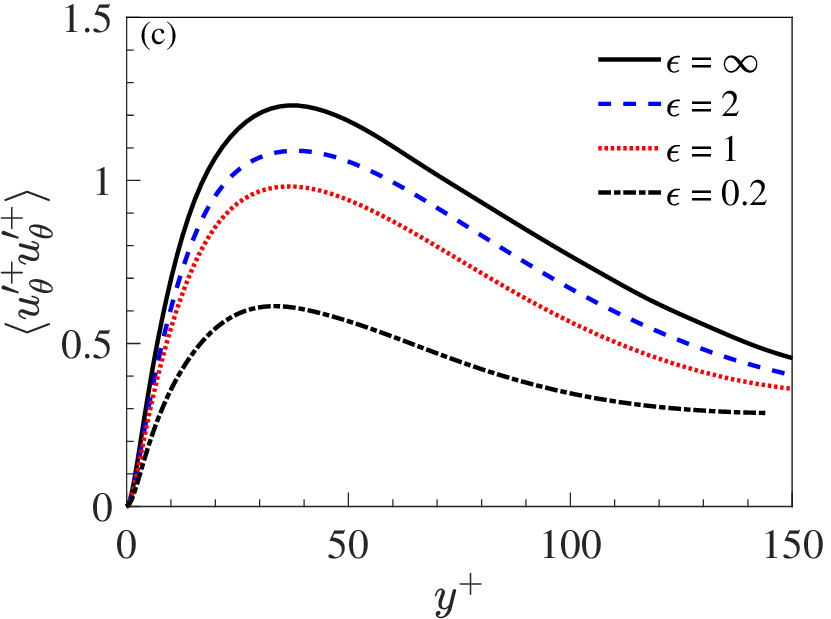}
    \end{subfigure}
    \begin{subfigure}{0.495\textwidth}
    \includegraphics[width=\linewidth]{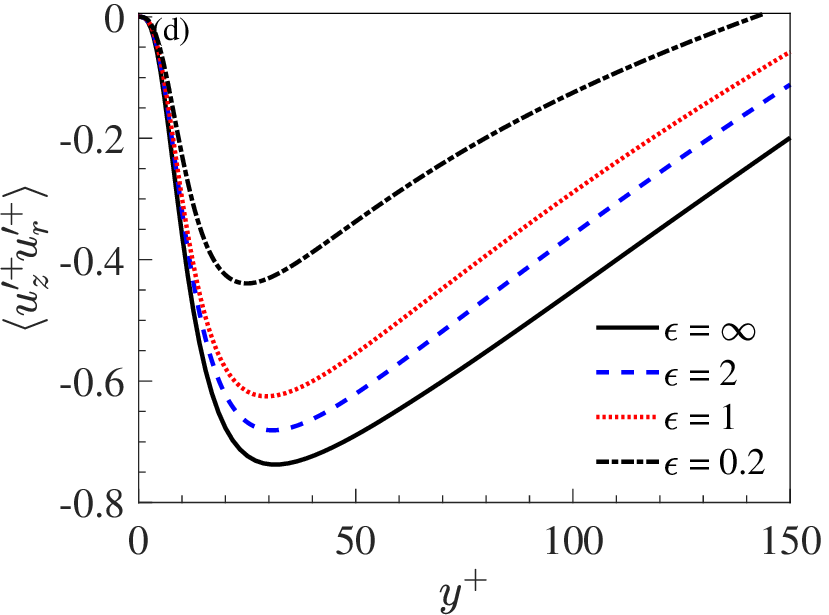}
    \end{subfigure}
    \caption{Wall normal distribution of the Reynolds stresses of (a)
    $\langle u^{\prime+}_zu^{\prime+}_z\rangle$, (b) $\langle
    u^{\prime+}_ru^{\prime+}_r\rangle$, (c) $\langle u^{\prime+}_\theta
    u^{\prime+}_\theta\rangle$ and (d) $\langle u^{\prime+}_z
    u^{\prime+}_r\rangle$ over inner cylinder wall.}
    \label{fig:ReynoldsInner}
\end{figure}

Figure~\ref{fig:ReynoldsInner}(b) shows the wall-normal distribution of $\langle
u^{\prime+}_ru^{\prime+}_r\rangle$. A striking difference is the amplitude that
is an order of magnitude lower compared to that of $\langle
u^{\prime+}_zu^{\prime+}_z\rangle$. This is well-known, implying that turbulent
energy is dominated by the streamwise velocity fluctuations. In addition, this
component peaks further away from the wall at $y^+=50$. Increasing curvature
appears to shift this peak location slightly closer to the wall, but reduces the
amplitude considerably. Such a decrease is evident even when $\epsilon=2$ and
$\epsilon=1$. In particular, when $\epsilon=0.2$, the profile shows a rather
flat behaviour beyond $y^+=50$. Figure~\ref{fig:ReynoldsInner}(c) shows
the wall-normal distribution of $\langle u^{\prime+}_\theta u^{\prime+}_\theta
\rangle$. The distribution of this component is very similar to that of $\langle
u^{\prime+}_ru^{\prime+}_r\rangle$, apart from the fact that $\langle
u^{\prime+}_\theta u^{\prime+}_\theta \rangle$ obtains slightly larger values.
This shows that the azimuthal fluctuations are stronger than the radial
fluctuations, a fact that does not appear to change with curvature.

\begin{figure}
    \centering
    \begin{subfigure}{0.495\textwidth}
    \includegraphics[width=\linewidth]{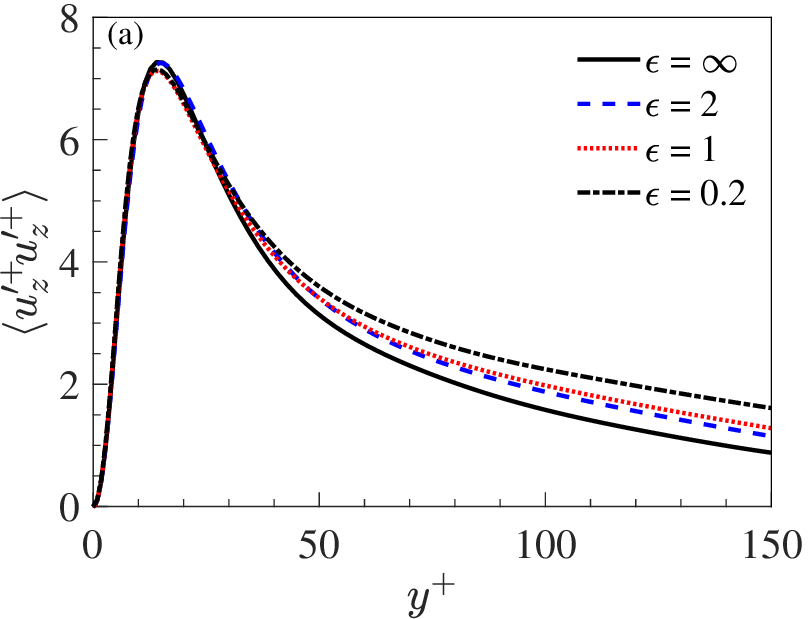}
    \end{subfigure}
    \begin{subfigure}{0.495\textwidth}
    \includegraphics[width=\linewidth]{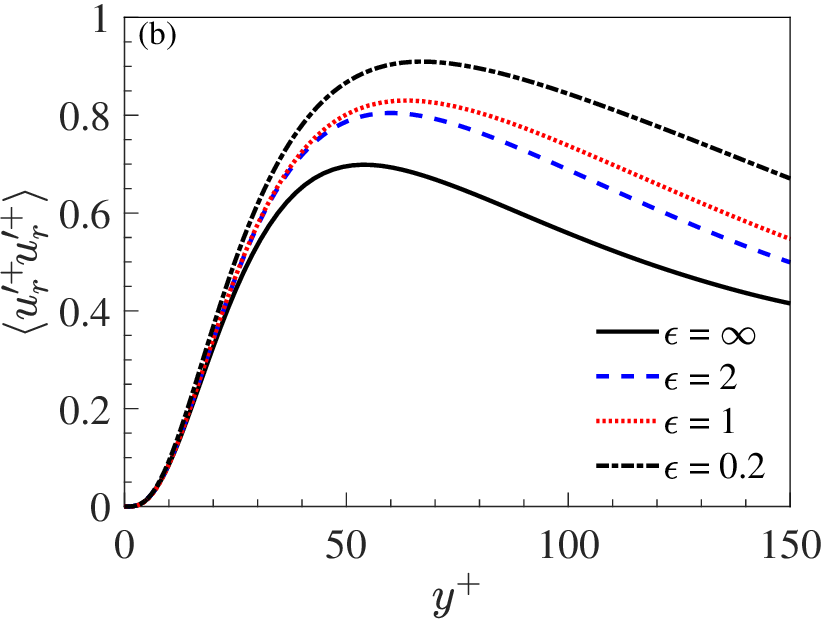}
    \end{subfigure}
    \begin{subfigure}{0.495\textwidth}
    \includegraphics[width=\linewidth]{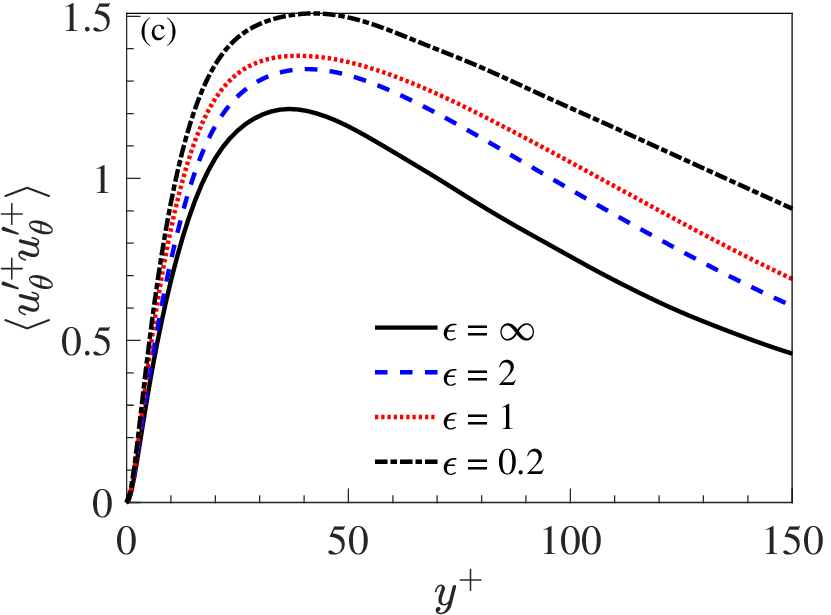}
    \end{subfigure}
    \begin{subfigure}{0.495\textwidth}
    \includegraphics[width=\linewidth]{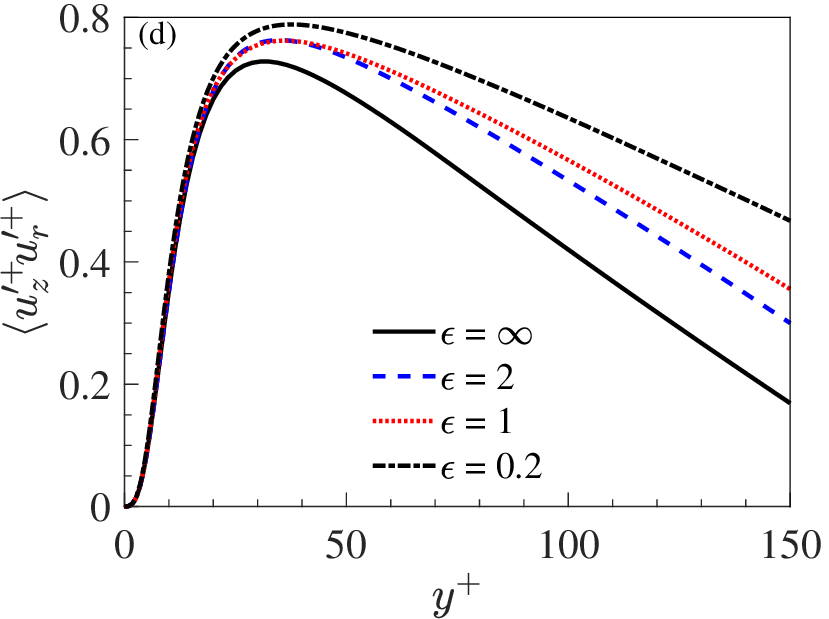}
    \end{subfigure}
    \caption{Wall normal distribution of the Reynolds stresses of (a) $\langle
    u^{\prime+}_zu^{\prime+}_z\rangle$, (b) $\langle
    u^{\prime+}_ru^{\prime+}_r\rangle$, (c) $\langle u^{\prime+}_\theta
    u^{\prime+}_\theta\rangle$ and (d) $\langle u^{\prime+}_z
    u^{\prime+}_r\rangle$ over outer cylinder wall.}
    \label{fig:ReynoldsOuter}
\end{figure}
Figure~\ref{fig:ReynoldsInner}(d) shows the wall-normal distribution of $\langle
u^{\prime+}_z u^{\prime+}_r \rangle$. Within the $y^+$ range shown in this
figure, $\langle u^{\prime+}_z u^{\prime+}_r\rangle$ is negative, indicating
that this is turbulent production term~\citep{Pope2000}. This
component peaks around $y^+=25$. As the curvature increases, its magnitude 
decreases considerably. Since the flow field is statistically homogeneous along
the $z$ and $\theta$ directions, the Reynolds stresses $\langle u^{\prime+}_z
u^{\prime+}_\theta \rangle$ and $\langle u^{\prime+}_r u^{\prime+}_\theta
\rangle$ are, as expected, approximately equal to $0$. We omit their
presentation in the text and only show them in Appendix A for reference.

Although we are interested in the WPF over the inner cylinder wall, it is
interesting to examine how the curvature changes the flow over the outer
cylinder wall. The curvature may be regarded as negative in this case. The
Reynolds stresses are shown in figure~\ref{fig:ReynoldsOuter}. One can see that
the effects of negative curvature on the flow are opposite to those of positive
curvature for all the Reynolds stress components, i.e. they increase as the
curvature increases. In addition, for all the stress components, their peak
positions move slightly away from the wall, as opposed to moving closer to the
wall for inner cylinders. Note that in figure~\ref{fig:ReynoldsOuter}(d), the
$\langle u^{\prime+}_r u^{\prime+}_z\rangle$ obtain positive values. This is
because the $u^{\prime+}_r$ represents the negative wall-normal velocity
component for the outer wall. As such, it still represents a term for
turbulence production. The $\langle u^{\prime+}_z u^{\prime+}_\theta \rangle$
and $\langle u^{\prime+}_r u^{\prime+}_\theta \rangle$ are again close to $0$,
details of which are also shown in appendix A.
}

\section{Results}
\label{sec:Results}
\add{Having validated the present DNS and discussed the effects of curvature on
the velocity statistics, we are now in a position to examine how the curvature
changes the characteristics of WPF. Note that the WPF} $p^\prime(z, \theta, t)$
is sampled with a time step $\Delta t U_b / \delta \approx 0.048$ for a total of
around $8000$ snapshots. The snapshot sequence is divided into a number of
segments with a $50\%$ overlap. \add{This results in 20 segments}, and each
segment contains 768 snapshots and is Fourier transformed/expanded along $z$,
$\theta$ and $t$ to yield its spectrum $\tilde{p}(k_z, m, \omega)$, i.e. 
\begin{equation}
  \tilde{p}(k_z, m, \omega) = \int_{0}^T \int_0^{2\pi} \int_{0}^{L_z} p^\prime(z, \theta, t) \e^{-\i \omega t} \e^{\i k_z z} \e^{\i m \theta}  \ud z \ud \theta \ud t.
\end{equation}
This yields a resolved non-dimensional frequency up to $\bar{\omega}\equiv\omega
\delta / U_b \approx 65$ with a resolution of around $0.165$ and a resolved
non-dimensional wavenumber up to $k_z \delta \approx 64$ with a resolution of
around $0.125$. 

The two-dimensional (2D) wavenumber-frequency spectrum is obtained from 
\add{
\begin{equation}
    \Phi(k_z, m, \omega) = \frac{1}{(2\pi)^2L_z T}\langle \tilde{p}(k_z, m,
    \omega) \tilde{p}^\ast(k_z, m, \omega) \rangle,
\end{equation}
} where $T$ represents the temporal duration of each segment. Note that each
segment is weighted by a Hanning window before the \add{Fourier transform
with respect to $t$} is performed. This procedure is pivotal to revealing the
low-wavenumber components simultaneously with the much more energetic convective
ridge by suppressing spectral leaks.

The one-dimensional (1D) wavenumber-frequency spectrum is calculated by summing
all azimuthal modes to yield
\add{
\begin{equation}
    \Phi(k_z, \omega) = \sum_{m=-\infty}^\infty \Phi(k_z, m, \omega).
\end{equation}

Similarly, the PSD of the WPF is obtained by
\begin{equation}
   \Phi(\omega) = \frac{1}{2\pi}\int_{-\infty}^\infty \Phi(k_z, \omega) \ud k_z. 
\end{equation}
} In the rest of this section, the PSD, 1D and 2D wavenumber-frequency spectra
at different $\epsilon$ values are examined sequentially. Note that the spectra
at $\epsilon=2$ show minimal difference from those at $\epsilon=\infty$; hence,
the $\epsilon=\infty$ case is omitted in the following comparison for
conciseness \add{and consistent definition of azimuthal modes (the
$\epsilon=\infty$ case needs a different definition for azimuthal modes).}

\subsection{Power spectral density}
\begin{figure}
    \centering
    \includegraphics[width=0.55\textwidth]{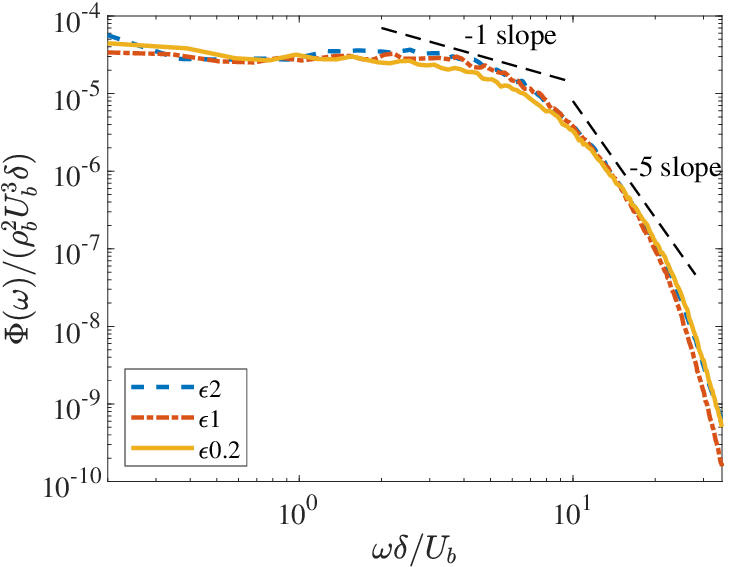}
    \caption{\add{Nondimensionalzed PSD $\Phi(\omega)/(\rho_b^2 U_b^3\delta)$
    of the WPF over cylinders of various diameters.}}
    \label{fig:PSD}
\end{figure}
Figure~\ref{fig:PSD} shows the comparison of the one-point temporal spectra of
the WPF over cylinders of various diameters. \add{All spectra exhibit a
$\omega^{-1}$ scaling in the intermediate but a $\omega^{-5}$ scaling in the
high frequency regimes. This is similar to the plane channel
flows~\citep{Liu2024, Bull1996}.} Clearly, the spectra of $\epsilon=2$ and
$\epsilon=1$ show minimal difference. This is well-known in incompressible
studies - when the inner cylinder is large, the effects of curvature on the WPF
are negligible~\citep{Neves1994b}. However, when $\epsilon$ reduces to $0.2$,
the wall pressure spectrum shows a clear reduction in magnitude in the
intermediate frequency regime. At high frequencies, however, the spectral
magnitude exhibits a slight increase compared to that at $\epsilon=1$. The
effects of curvature on the low-frequency PSD component appear rather limited.
Note, however, this does not contradict the findings of \citet{Neves1994b},
where the PSD reduces at all frequencies when the friction velocity $u_\tau$ is
used for \add{non-dimensionalisation.} To examine the structures of the WPF,
however, one is more interested in the wavenumber-frequency spectra shown in the
following sections.


\subsection{1D wavenumber-frequency spectra}
\begin{figure}
   \includegraphics[width=0.95\textwidth]{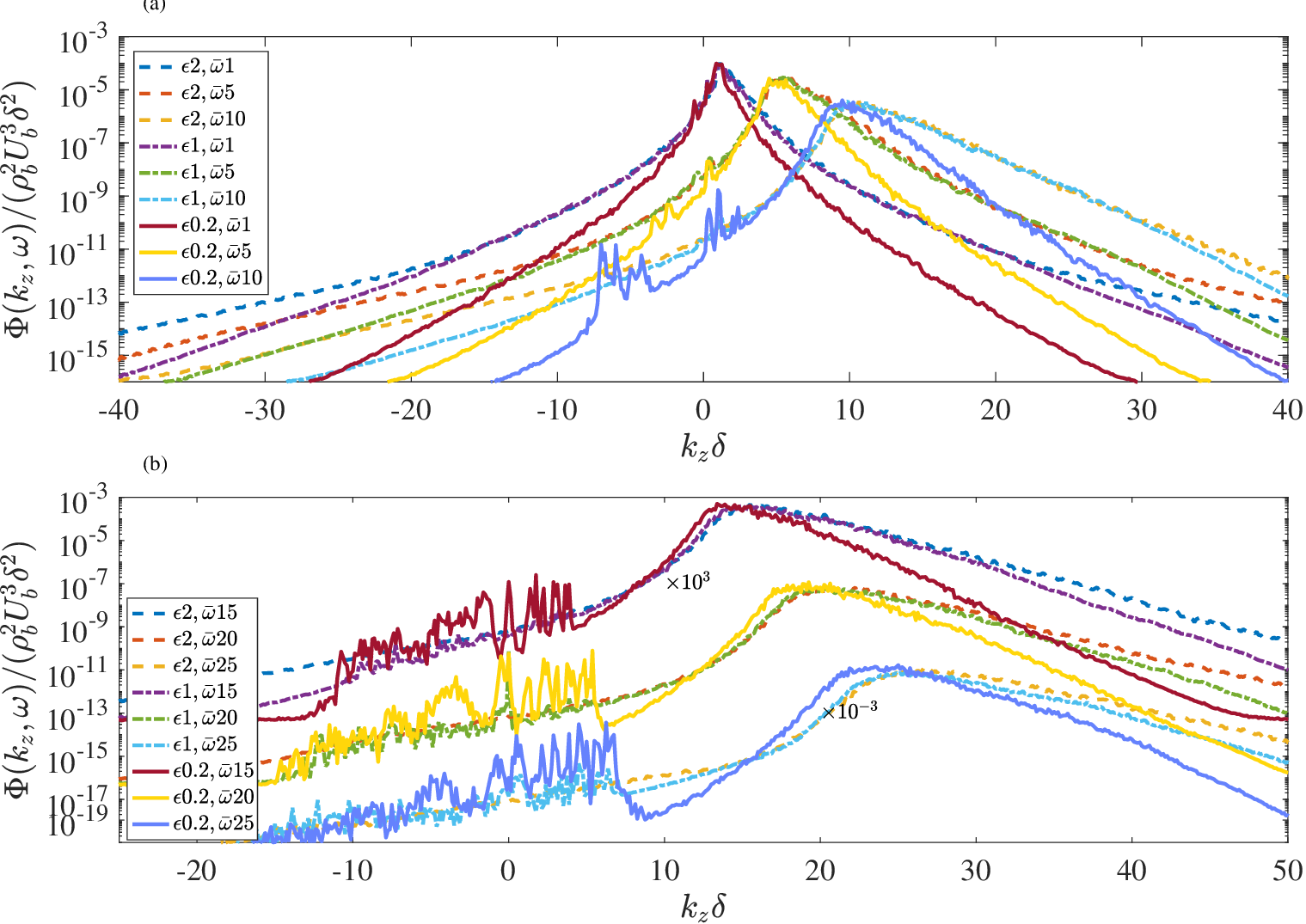}
   \caption{\add{1D wavenumber-frequency $\Phi(k_z, \omega)/(\rho_b^2 U_b^3
   \delta^2)$ spectra of WPF for various cylinder diameters: (a)
   $\bar{\omega}=1$, $5$ and $10$; (b) $\bar{\omega}=15$, $20$ and $25$ the
   spectral magnitudes at $\bar{\omega}=15$ and $\bar{\omega}=25$ are multiplied
   by $10^3$ and $10^{-3}$ respectively for separation clarity.}}
   \label{fig:0n1DSpectra}
\end{figure}

Figure~\ref{fig:0n1DSpectra}(b-c) compares the 1D wavenumber-frequency spectra
$\Phi(k_z, \omega)$ when $\epsilon$ varies. At a low frequency of
$\bar{\omega}=1$, one can see a clear dominant peak at the \add{convective}
wavenumbers $\omega/U_c$. Reducing the radius from $\epsilon=2$ to $1$ lowers
the magnitude at high (including negative) wavenumbers. Further decreasing the
cylinder radius to $\epsilon=0.2$ reduces the spectra more significantly.
However, the magnitude at the \add{convective} wavenumber appears virtually
unchanged. This implies that the streamwise length scale of the WPF becomes
larger, a result in accordance with earlier incompressible
studies~\citep{Neves1994b}. In addition, one starts to see a significant
spectral feature change when $\epsilon$ reduces to $0.2$, i.e. acoustic peaks
(see section~\ref{subsec:2D}) start to appear in the supersonic wavenumber
region. 

As the frequency increases to $\bar{\omega} = 5$ and $10$, the convection peak moves to higher wavenumbers with reduced magnitudes, consistent with a decaying magnitude against frequency shown in figure~\ref{fig:0n1DSpectra}(a). The effects of reducing $\epsilon$ are similar to those at $\bar{\omega} = 1$, except that the acoustic peaks at $\epsilon=0.2$ are increasingly strong as the frequency increases. In addition, the spectra appear to shift to lower wavenumbers as $R$ reduces. 


The acoustic peaks and spectral shift are more significant at higher
frequencies. Figure~\ref{fig:0n1DSpectra}(c) shows the spectra at
$\bar{\omega}=15$, $20$ and $25$, respectively. At $\bar{\omega}=15$, the
spectrum is smooth at all wavenumbers at $\epsilon=2$ but starts to show some
acoustic peaks at $\epsilon = 1$. At $\epsilon=0.2$, a much denser cluster of
acoustic peaks starts to occur. As the frequency increases to $\bar{\omega}=20$
and $25$, these acoustic peaks are both stronger and denser. \add{Note that
acoustic peaks do not appear in the case of rectangular channels~\citep{Liu2024}
and are caused by the curvature of the walls.} Similar to the appearance of the
acoustic peaks, the tendency of spectral shift appears increasingly evident as
the frequency increases.

\subsection{2D wavenumber-frequency spectra}
\label{subsec:2D}
\begin{figure}
    \centering
    \includegraphics[width=0.98\textwidth]{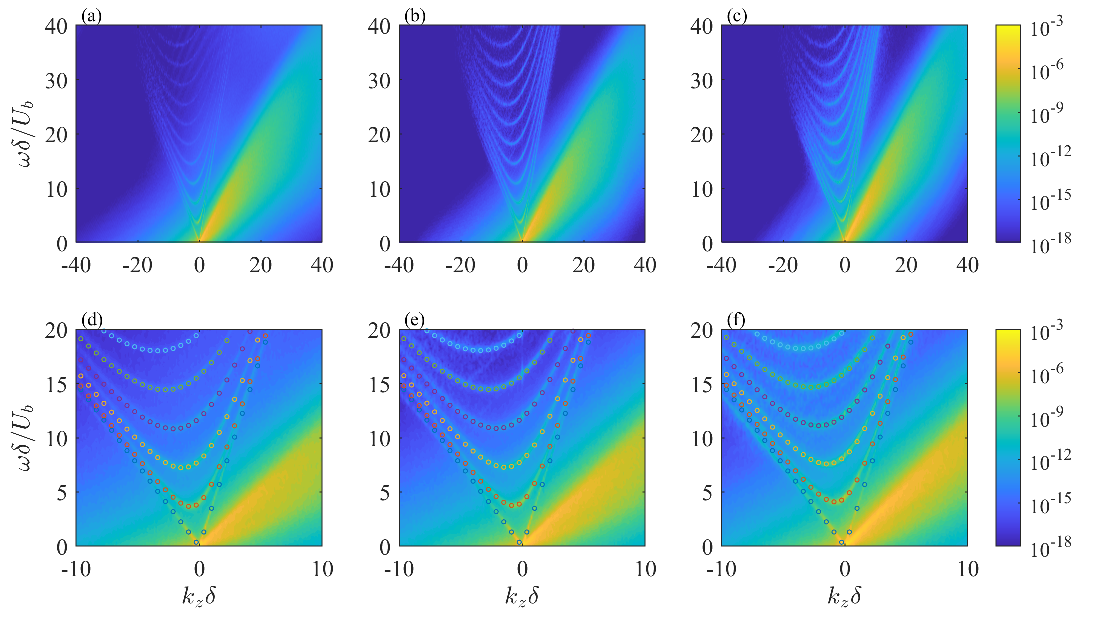}
    \caption{The 2D wavenumber-frequency spectra at $\epsilon=2$ (a,d), $1$
    (b,c) and $0.2$ (e,f). The dotted lines represent the acoustic duct modes
    solved from (\ref{equ:crossEquation}).}
    \label{fig:2DContour}
\end{figure}
The fact that the spectral peaks observed in figure~\ref{fig:0n1DSpectra}(c) are acoustics induced can be clearly seen from the 2D wavenumber-frequency spectra $\Phi(k_z, m, \omega)$. Considering that it is the axisymmetric mode ($m=0$) that is more likely to couple with the elastic structures, figure~\ref{fig:2DContour} first shows the contour of $\Phi(k_z, 0, \omega)$ over cylinders of various diameters. Note that the colour is plotted in a logarithmic scale due to the wide span of the spectral magnitude.

Figure~\ref{fig:2DContour}(a) shows $\Phi(k_z, 0, \omega)$ for $\epsilon=2$. Due
to compressibility, we see an evident violation of the Kraichnan-Phillips
theorem that demands $\Phi(k_z, k_y, \omega) \to 0$ as $k_z^2+k_y^2 \to
0$~\citep{williams_surface-pressure_1965}, where $k_y$ represents the spanwise
wavenumber. Similar to the incompressible case~\citep{Neves1994b}, the energy of
the WPF is dominated by the convective ridges. As the frequency increases,
however, these ridges become increasingly wide, suggesting a less well-defined
convection velocity. What sets the compressible spectra apart from their
incompressible counterparts, however, are a series of discrete curves clearly
visible albeit at a much weaker intensity. \add{These are the acoustic modes in
annular ducts, which are similar to those in plane channel flows~\citep{Liu2024}.}

To examine these acoustic modes closely, figure~\ref{fig:2DContour}(d-f)
shows a close-up view of the low-wavenumber regimes of
figure~\ref{fig:2DContour}(a-c). Also included are the analytical dispersion
relations of the acoustic modes in annular ducts with uniform mean
flow~\citep{crighton_modern_1992}, i.e.
\begin{equation}
    (\beta^2 k_z + k M_b)^2 = k^2 - \mu_{mn}^2\beta^2,
\label{equ:crossEquation}
\end{equation}
where $\beta^2 = 1-M_b^2$, $k=\omega / c_0$, while $\mu_{mn}$ is the $n$th root
of 
\add{
\begin{equation}
    J_m^\prime(\mu_{mn} R) Y_m^\prime(\mu_{mn} (R+2\delta)) -
J_m^\prime(\mu_{mn} (R+2\delta)) Y_m^\prime(\mu_{mn}R) = 0.
\label{equ:dispersion}
\end{equation}
} In (\ref{equ:dispersion}), $J_m$ and $Y_m$ represent the $m$th-order Bessel
functions of the first and second kinds, respectively, \add{the prime
superscript represents their first derivatives, while $n$ denotes an integer
representing the radial order of the acoustic mode.} One can see the nearly
perfect collapse of these predicted duct modes with those discrete curves in
figure~\ref{fig:2DContour}(d-f). Note the striking difference in the spectral
magnitudes of these acoustic modes and the convective ridge. However, as
$\epsilon$ reduces, these acoustic modes appear to intensify, as shown by the
increasingly bright lines in figure~\ref{fig:2DContour}(b-f).

\add{To examine the acoustic modes in detail, we plot their pressure
distribution within the annular ducts of various diameters. These acoustic modes
can be calculated analytically, and they take the form of
\begin{equation}
    p=C_{mn}\left(Y_m^\prime(\mu_{mn}R) J_m(\mu_{mn}r)-
    J_m^\prime(\mu_{mn}R)Y_m(\mu_{mn}\ r)\right) 
\exp(im\theta)\exp{(-ik_z z)},
\label{equ:acousticMode}
\end{equation}
where, $C_{mn}$ represents the normalisation constant that depends on the
azimuthal order $m$ and radial order $n$. In this paper, we focus on the
axisymmetric mode $m=0$ and use (\ref{equ:acousticMode}) to plot the acoustic
mode propagating toward the $+z$ direction. Those propagating towards the $-z$
direction are similar and hence omitted.

\begin{figure}
    \centering
    \includegraphics[width=0.9\textwidth]{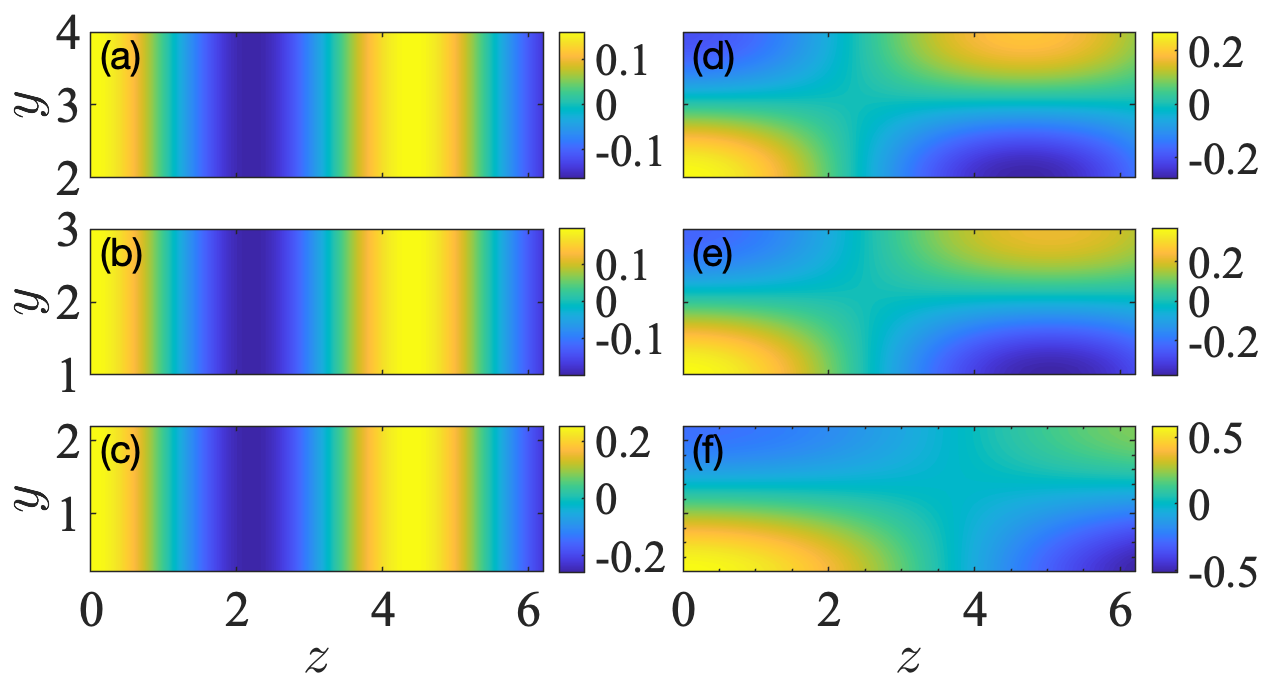}
    \caption{\add{Axisymmetric ($m=0$) acoustic modes at $k=2$: (a-c) modes of
    radial order $n=1$ (plane waves); (d-f) modes of radial order $n=2$; (a,d):
    $\epsilon=2$; (b,e) $\epsilon=1$; (c, f) $\epsilon=0.2$.}}
    \label{fig:AcousticModeK2}
\end{figure}

\begin{figure}
    \centering
    \includegraphics[width=0.9\textwidth]{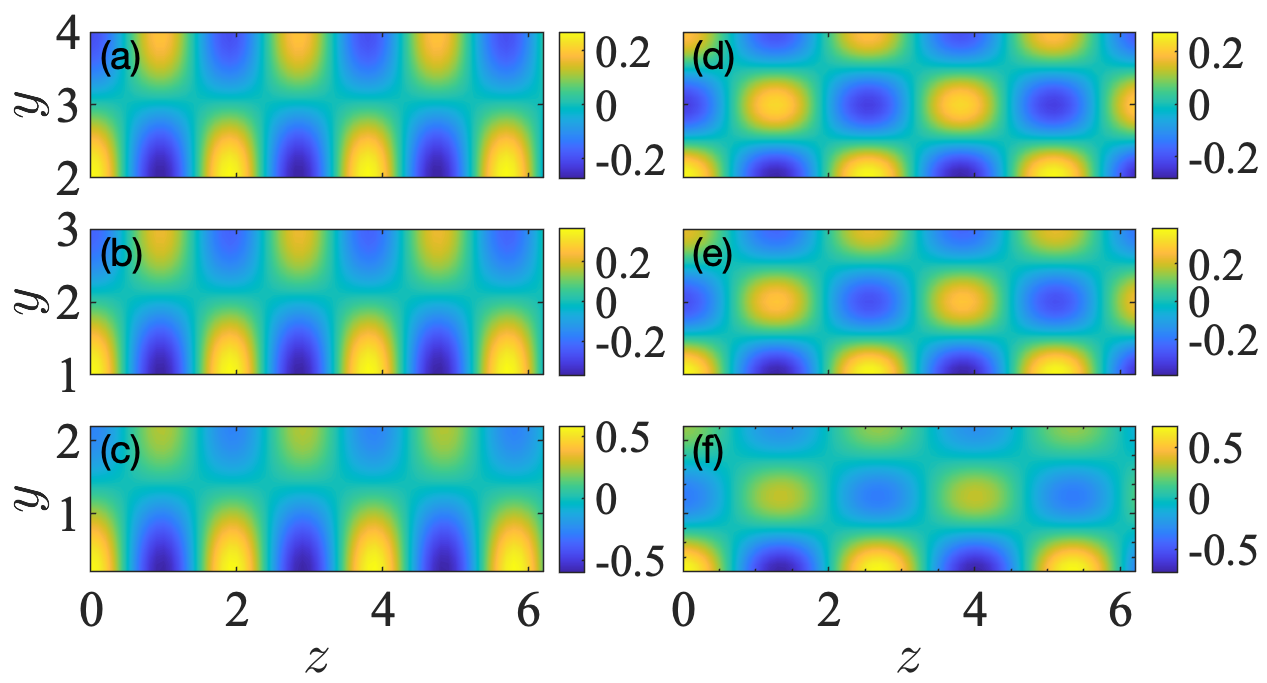}
    \caption{\add{Axisymmetric ($m=0$) acoustic modes at $k=5$: (a-c) modes of
    radial order $n=2$; (d-f) modes of radial order $n=3$; (a,d): $\epsilon=2$;
    (b,e) $\epsilon=1$; (c, f) $\epsilon=0.2$.}}
    \label{fig:AcousticModeK5}
\end{figure}

The cross-sections of the duct modes for the three cylinder diameters are shown in figures~\ref{fig:AcousticModeK2} and \ref{fig:AcousticModeK5} at a reduced frequency of $k\delta=2$ and $5$, respectively. Figure~\ref{fig:AcousticModeK2}(a-c) shows the real part of (\ref{equ:acousticMode}) for the acoustic mode of radial order $n=1$ at $k\delta=2$. These represent plane wave modes (the straight discrete lines in figure~\ref{fig:2DContour}) propagating along the $+z$ direction. In figure~\ref{fig:2DContour}, we show that these plane-wave modes appear stronger on the inner cylinder wall as $\epsilon$ reduces. Since their amplitudes on the inner and outer cylinder walls remains identical, as shown in figure~\ref{fig:AcousticModeK2}(a-c), one expects a similar behaviour on the outer wall. 

Figure~\ref{fig:AcousticModeK2}(d-f), on the other hand, shows the acoustic
modes of the radial order $n=2$. It worth noting that only two radial orders
($1$ and $2$) are cut-on at this reduced frequency of $k\delta=2$, while
higher-order modes are effectively cut-off. This can be seen clearly from
figure~\ref{fig:2DContour}(d-f) by drawing a horizontal line at
$\omega\delta/U_b=5$. From figure~\ref{fig:AcousticModeK2}(d-f), we can see that
as the curvature increases, the acoustic pressure on the inner cylinder wall
becomes increasingly larger than that on the outer wall. To put this difference
into perspective, the acoustic pressure at the inner wall is approximately 10 dB
higher than that on the outer wall. One therefore expects less energetic
acoustic peaks for the radial order $n=2$ on the outer wall.

As the frequency increases, figure~\ref{fig:2DContour}(d-f) shows that more
acoustic modes of higher radial orders become cut-on. To show these modes more
closely, we choose a higher frequency of $k\delta=10$ and plot the acoustic duct
modes of radial orders $2$ and $3$ in figure~\ref{fig:AcousticModeK5}. Again,
high-order modes are cut-off. As the radial order increases, the acoustic
pressure shows more zeros along the radial direction, while the streamwise
wavenumber decreases. The latter is expected because of (\ref{equ:dispersion}).
Moreover, the difference in the pressure amplitude on the inner and outer walls
appears larger as $n$ increases. This suggests that the higher-order acoustic
peaks in the WPF spectra are increasingly more pronounced on the inner wall than
those on the outer wall. Note, however, although we can examine the behaviour of
WPF on the outer walls, our primary concern in practice, and hence in this paper,
is the WPF over the inner cylinder walls.}

\begin{figure}
    \includegraphics[width=0.95\textwidth]{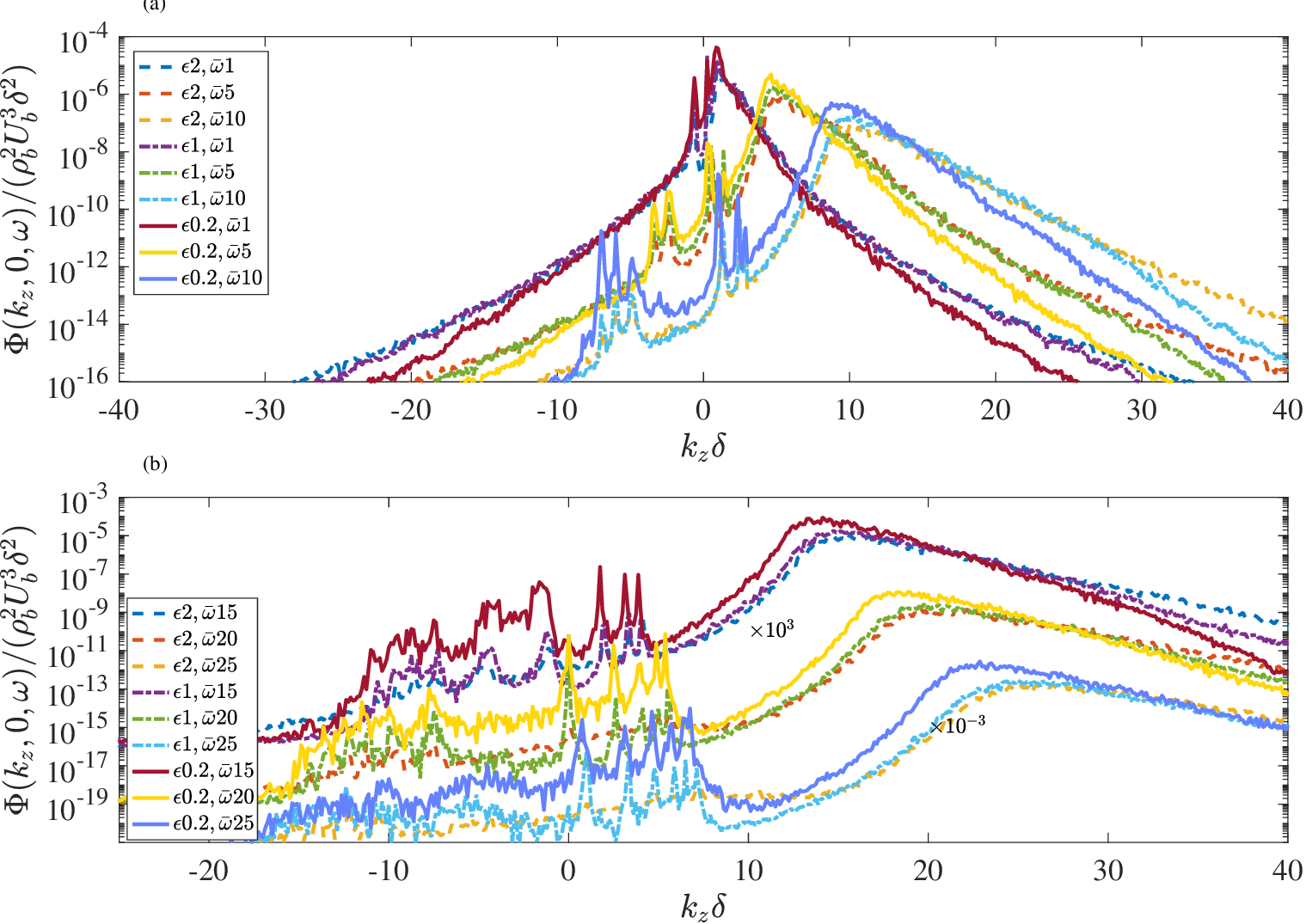}
    \caption{The non-dimensional 2D wavenumber-frequency spectra $\Phi(k_z, 0,
    \omega)/(\rho_b^2 U_b^3 \delta^2)$ over cylinders of various diameters; the
    spectral magnitudes at $\bar{\omega}=15$ and $\bar{\omega}=25$ are
    multiplied by $10^3$ and $10^{-3}$ respectively for separation clarity. }
    \label{fig:2Dspectra}
\end{figure}
To compare the 2D wavenumber-frequency spectra over inner walls quantitatively, figure~\ref{fig:2Dspectra}(a) shows $\Phi_{pp}(k_z, 0, \omega)$ against $k_z\delta$ at $\bar{\omega} = 1$, $5$ and $10$. One can see that decreasing the diameter increases the spectral magnitude at low but decreases it at high wavenumbers. 
Different from figure~\ref{fig:0n1DSpectra}(b), spectral peaks due to acoustic duct modes can be found at all cylinder diameters. As shown in figure~\ref{fig:2DContour}, these peak wavenumbers can be well predicted by (\ref{equ:crossEquation}). As the frequency increases, the low-wavenumber spectral augmentation appears stronger, as shown in figure~\ref{fig:2Dspectra}(b). \add{Note that when $\bar{\omega} \ge 20$, the low-wavenumber spectral magnitude at $\epsilon=2$ is broadly elevated without apparent peaks, in accordance with \citet{Liu2024}.} But the acoustic peaks continue to show up when $\epsilon=1$ and $0.2$. In addition, the acoustic peaks become more clustered because more duct modes are cut-on, as shown straightforwardly in figure~\ref{fig:2DContour}. Clearly, when all azimuthal modes are summed to obtain $\Phi(k_z, \omega)$, more spectral peaks will appear, as evidenced by figure~\ref{fig:0n1DSpectra}.

\begin{figure}
	\includegraphics[width=0.95\textwidth]{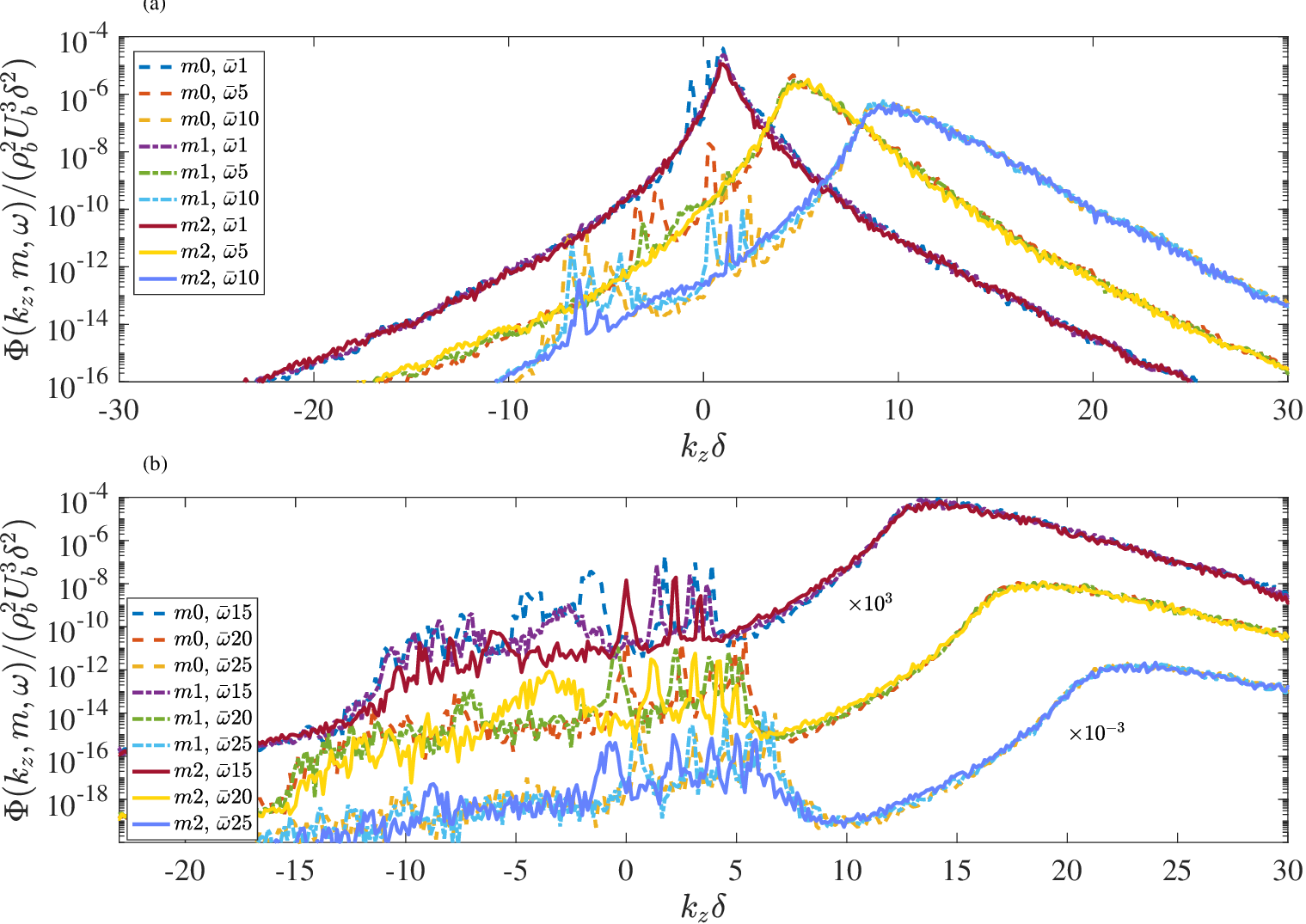}
	\caption{The non-dimensional 2D wavenumber-frequency spectra $\Phi(k_z,
	m, \omega)/(\rho_b^2 U_b^3 \delta^2)$ for azimuthal modes $m=0$, $1$ and
	$2$ for $\epsilon = 0.2$; the spectral magnitudes at $\bar{\omega}=15$
	and $\bar{\omega}=25$ are multiplied by $10^3$ and $10^{-3}$
	respectively for separation clarity.}
    \label{fig:mIncrease}
\end{figure}
To show the variation of the wavenumber-frequency spectra against azimuthal
modes, figure~\ref{fig:mIncrease} shows the spectra at various $m$ for
$\epsilon=0.2$. When the frequency is low ($\bar{\omega}=1$), one can see that
only the spectra at $m=0$ exhibit acoustic peaks, since high-order modes are
effectively cut-off according to (\ref{equ:crossEquation}). The spectral
magnitude outside the acoustic peaks remains roughly identical. As the frequency
increases to 5 and 10, acoustic modes at $m=1$ and $m=2$ become respectively
cut-on. As the frequency further increases, more acoustic modes are cut-on and
appear in the spectra shown in figure~\ref{fig:mIncrease}(b). Note that the
elevated magnitude in the supersonic region appears to exist for all $m$. Other
than that, the spectra look similar elsewhere. This suggests a possible
necessity of including WPF of higher azimuthal orders in some flow noise
applications.


\section{Discussions}
\label{sec:discussions}
\label{subsec:WPFmodel}
\add{The low-wavenumber augmentation in spectral magnitude as $\epsilon$ reduces
is of great concern in practical applications such as towed sonar
arrays~\citep{Bull1996}. 
To understand why such augmentation occurs, we start from modelling the WPF
generated by a homogeneous turbulent boundary layer over an infinitely long
cylinder. Note that such a model is not strictly applicable to the present
setup, where channel flows within concentric annular ducts are simulated instead
of free boundary layers. Nevertheless, we expect the effects of curvature on the
WPF in these two cases remains similar, particularly at low wavenumbers.
Therefore, we attempt to discuss the low-wavenumber WPF augmentation via
modelling the boundary layer case. }

\subsection{Modelling WPF based on the Lighthill acoustic analogy}
Using Lighthill's acoustic analogy and the acoustic Green's function, the WPF
spectrum beneath a homogeneous turbulent boundary layer over an infinitely long
cylinder can be found \add{analytically} as~\citep{Dhanak1988}
\begin{equation}
    \tilde{p}(k_z, m, \omega)
    = \int T_{ij}(\boldsymbol{y}, \tau)
    \frac{\partial^2 }{\partial y_i \partial y_j}
    \left(
    -\frac{1}{\alpha R}
    \frac{H_m^{(1)}(\alpha r)}{H_m^{(1)\prime}(\alpha R)}
    \e^{\i k_z y }
    \e^{\i m \phi}
    \e^{-\i \omega\tau}
    \right)
    \ud^3 \boldsymbol{y}
    \ud \tau,  
    \label{equ:Dhanak}
\end{equation} 
where $T_{ij}$ ($i,j=1,2,3$) is the \add{Cartesian} Lighthill's stress tensor,
$\alpha = \sqrt{\omega^2/c_0^2 - k_z^2}$, $H_m^{(1)}(z)$ is the $m$th-order
Hankel function of the first kind.  The spatial integration in
(\ref{equ:Dhanak}) is performed within the entire boundary layer \add{over the
cylinder}. The 0th-order mode is of particular relevance, \add{and in
cylindrical coordinates shown in figure~\ref{fig:case_config}} it reduces to
\add{
\begin{IEEEeqnarray}{rCl}
    \int_R^{R+\delta}
    \left[
	\frac{-\alpha H_0^{(1)\prime\prime}(\alpha r)}
	{H_0^{(1)\prime}(\alpha R)}\tilde{T}_{r r}
	-\frac{H_0^{(1)\prime} (\alpha r)}{H_0^{(1)\prime}(\alpha R)}
	\left( \frac{\tilde{T}_{\theta\theta}}{r}
     + 2\i k_z \tilde{T}_{r z}\right) \right.
     \left. 
	+ \frac{k_z^2 H_0^{(1)}(\alpha r)}{\alpha H_0^{(1)\prime}(\alpha R)}
	\tilde{T}_{zz}
    \right] 
    \frac{r}{R}
    \ud r, \IEEEeqnarraynumspace 
    \label{equ:0mode}
\end{IEEEeqnarray}
where the integral along the radial direction is from the cylinder radius
$R$ to the edge of the boundary layer $R+\delta$, while $\tilde{T}_{ij}$ is the
Fourier transform of Lighthill's stress tensor given by 
\begin{equation}
    \tilde{T}_{ij}(r, k_z, 0, \omega) = \iint T_{ij}(r, z, \theta,
\omega) \e^{\i k_z z} \e^{-\i \omega \tau}\ud z \ud \theta \ud \tau,
\end{equation}
where the subscripts $i$ and $j$ take the value of $r$, $\theta$ and $z$ in the
cylindrical coordinates. Equation (\ref{equ:0mode}) shows that the $0$th-order
WPF are only generated by the stress terms $\tilde{T}_{rr}$,
$\tilde{T}_{\theta\theta}$, $\tilde{T}_{r z}$ and $\tilde{T}_{zz}$ within the
boundary layer and are independent of $\tilde{T}_{\theta r}$ and
$\tilde{T}_{z\theta}$. The latter two components are also expected to be small
(see, for example, Appendix A for more details).}

In the vicinity of the sonic peak $k_z = \omega / c_0$, the last term in
(\ref{equ:0mode}) is singular, but integrable. In practice, the height of these
sonic peaks is regularised by the finite size and mean-flow refraction effects
of the boundary layer~\citep{Bull1996}, details of which are not discussed here
(note that in the channel flow case, more spectral peaks are expected apart from
the sonic peak because of the presence of multiple cut-on duct modes). In most
applications where the low-wavenumber components are of particular concern, one
often has $k_z r \ll 1$ ($r$ takes the value between $R$ and $R+\delta$). In
such cases, the last two terms of the right-hand side of (\ref{equ:0mode}) can
be neglected except near the sonic singularity. 

When the cylinder radius and boundary layer thickness are compact ($kR \ll 1$
and $k\delta\ll 1$, e.g. at low frequencies such as those in towed sonar
arrays), (\ref{equ:0mode}) can be approximated using the small-argument
expansion of the Hankel functions~\citep{crighton_modern_1992} as
\begin{IEEEeqnarray}{rCl} 
    \tilde{p}(k_z, 0, \omega) \approx &&\int_0^1\frac{\tilde{T}_{r r} -
    \tilde{T}_{\theta\theta} }{\epsilon+h} \ud h,
\label{equ:0modeCompact} 
\end{IEEEeqnarray}
\add{where $h\equiv y/\delta$ represents the non-dimensionalised radial distance
measured from the inner cylinder wall and therefore takes the value from $0$ to
$1$. On the other hand, when the cylinder radius is sufficiently non-compact
(e.g. at sufficiently high frequencies)}, i.e. $kR \gg 1$, the Hankel functions
can be expanded using their asymptotic form and (\ref{equ:0mode}) at low
wavenumbers can be approximated as
\begin{IEEEeqnarray}{rCl} 
    \tilde{p}(k_z, 0, \omega) \approx &&\int_0^1
    \sqrt{1+\frac{h}{\epsilon}} \alpha\delta \e^{\i \alpha \delta h} \e^{-\i \frac{\pi}{2}}
    \tilde{T}_{rr}
    -
    \frac{1}{\sqrt{\epsilon(\epsilon+h)}} \e^{\i \alpha \delta h }
    \tilde{T}_{\theta\theta}
     \ud h.
\label{equ:0modeNoncompact} 
\end{IEEEeqnarray}

Assuming that the distributions of Lighthill's stress tensors $\tilde{T}_{rr}$
and $\tilde{T}_{\theta\theta}$ between the lower ($h=0$) and upper walls ($h=2$)
do not change significantly as $R$ varies, (\ref{equ:0modeCompact}) and
(\ref{equ:0modeNoncompact}) show that the $0$th-order spectrum at small $k_z$
would increase when the radius of the inner cylinder reduces. This complies with
the low-wavenumber increase of $\Phi(k_z, 0, \omega)$ observed at both low and
high frequencies in figure~\ref{fig:2Dspectra}. \add{Note that the $\epsilon$
terms in equations (\ref{equ:0modeCompact}) and (\ref{equ:0modeNoncompact})
result from the asymptotic approximation of Hankel functions rather than from
the Lighthill's stress. This suggests that the low-wavenumber increase is a
\add{``geometric'' effect} connected with the Green's function.}

In particular, the appearance of $1/\sqrt{\epsilon(\epsilon+h)}$ in
(\ref{equ:0modeNoncompact}) as opposed to $1/(\epsilon+h)$ in
(\ref{equ:0modeCompact}) suggests that the amplification at high frequencies
\add{may be more pronounced than that at low frequencies. There exists indeed
such a tendency in figure~\ref{fig:2Dspectra} as the frequency increases.} Note
that only radial and azimuthal velocity disturbances appear important. Moreover,
both (\ref{equ:0modeCompact}) and (\ref{equ:0modeNoncompact}) suggest that as
the curvature increases, disturbances close to the wall are increasingly more
important in generating low-wavenumber components of the WPF spectrum. This
complies with the fact that the convective velocity at a fixed $\omega\delta /
u_\tau$ decreases as $R$ reduces. This signals a potential for low-wavenumber
WPF control using near-wall treatment on thin cylinders. 

\add{
\subsection{Comparison with DNS}
In the derivation reaching (\ref{equ:0modeCompact}) and
(\ref{equ:0modeNoncompact}), the assumption that the Lighthill stress tensors
$\tilde{T}_{rr}$ and $\tilde{T}_{\theta\theta}$ between the cylinder wall
($h=0$) and outer edge of the boundary layer ($h=1$) do not change significantly
as $R$ varies is important. To examine to what extent this assumption remains
valid, we first show the wavenumber-frequency spectra of the two stress
components, i.e. $\Psi_{r r}(r, k_z, 0, \omega)$ and $\Psi_{\theta\theta}(r,
k_z, 0, \omega)$, along the wall-normal direction at a fixed frequency $\omega$
and wavenumber $k_z$. Note that the wavenumber-frequency spectra $\Psi_{rr}
\propto |\tilde{T}_{rr}|^2$ while $\Psi_{\theta\theta} \propto
|\tilde{T}_{\theta\theta}|^2$, therefore, $\sqrt{\Psi_{rr}}$ and
$\sqrt{\Psi_{\theta\theta}}$ are shown in figures~\ref{fig:TijInnerKxInd1f0} to
\ref{fig:TijInnerKxInd1f0p2}.

As mentioned above, we focus on the low-wavenumber region, where $k_z r \ll 1$
($r$ takes the value between $R$ to $R+\delta$). However, the streamwise
wavenumber resolution of the present simulation is only around $k_z \delta
\approx 0.125$. This implies that the minimum non-zero wavenumber that can be
resolved corresponds to a value of $k_z r$ up to $0.6$. Therefore, we only show
$\sqrt{\Psi_{r r}}$ and $\sqrt{\Psi_{\theta\theta}}$ at $k_z\delta =0$ in this
paper. Nevertheless, doing this does not reduce much generality, because the
wavenumber-frequency spectrum has a well-known wavenumber-white
behaviour~\citep{Bull1996} near $k_z\delta=0$, i.e. $\Phi(k_z, m, \omega)$
obtains a nearly constant value in the low-wavenumber range centring around
$k_z\delta=0$.

Note that the asymptotic form of (\ref{equ:0modeCompact}) requires the cylinder
to be acoustically compact, i.e. $kR \ll 1$ and $k\delta\ll 1$. This corresponds
to a low-frequency range, which we can properly resolve with a frequency
resolution of $\omega\delta / U_b\approx 0.165$ ($k\delta\approx 0.01$).
However, the asymptotic form of (\ref{equ:0modeNoncompact}) requires acoustic
non-compactness, i.e. $kR \gg 1$. This implies a considerably high frequency
range. To put this into perspective, for $\epsilon=0.2$, a moderate number
$kR=2$ corresponds to a frequency of $\omega \delta/ U_b = 157$. Resolving such
a frequency needs a very high sampling rate of the flow snapshots, and in
practice storing this tremendous amount of data is prohibitively expensive. More
importantly, in practical applications, it is the low frequencies that are of
our primary concern. As such, in the following, we focus on examining the
distribution of $\sqrt{\Psi_{r r}}$ and $\sqrt{\Psi_{\theta\theta}}$ at low
frequencies where the cylinder is acoustically compact.
\begin{figure} \centering
    \begin{subfigure}{0.495\textwidth}
    \includegraphics[width=\linewidth]{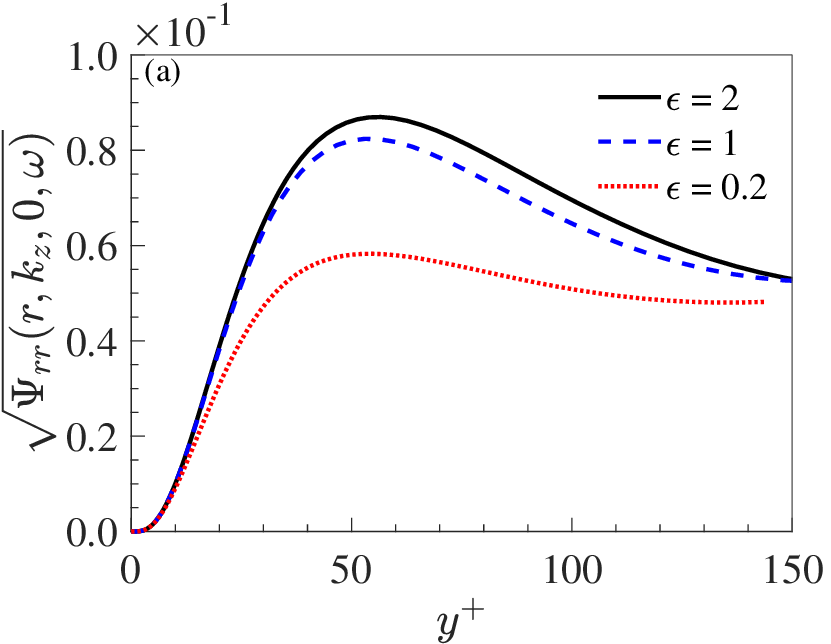}
    \end{subfigure}
    \begin{subfigure}{0.495\textwidth}
    \includegraphics[width=\linewidth]{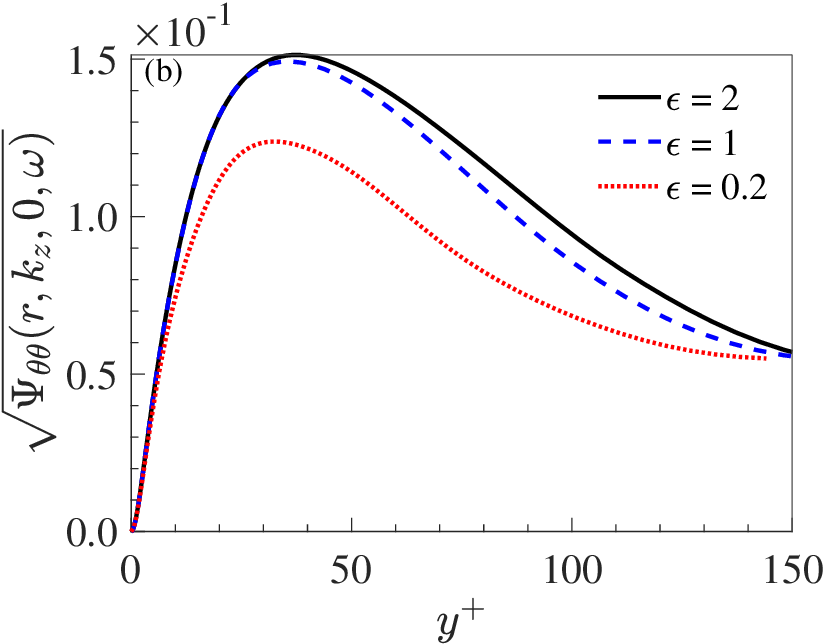}
    \end{subfigure}
    \caption{Wall normal distribution of the Lighthill stress components (a)
    $\tilde{T}_{rr}$ and (b) $\tilde{T}_{\theta\theta}$ over inner cylinder
    surface at $k_z\delta=0$ and $\omega\delta/U_b=0$.}
    \label{fig:TijInnerKxInd1f0}
\end{figure}
\begin{figure}
    \centering
    \begin{subfigure}{0.495\textwidth}
    \includegraphics[width=\linewidth]{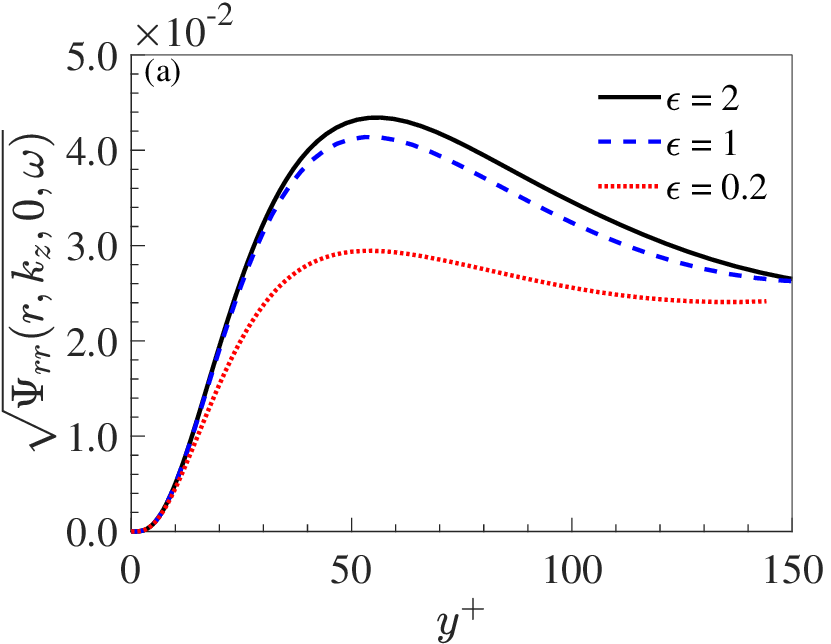}
    \end{subfigure}
    \begin{subfigure}{0.495\textwidth}
    \includegraphics[width=\linewidth]{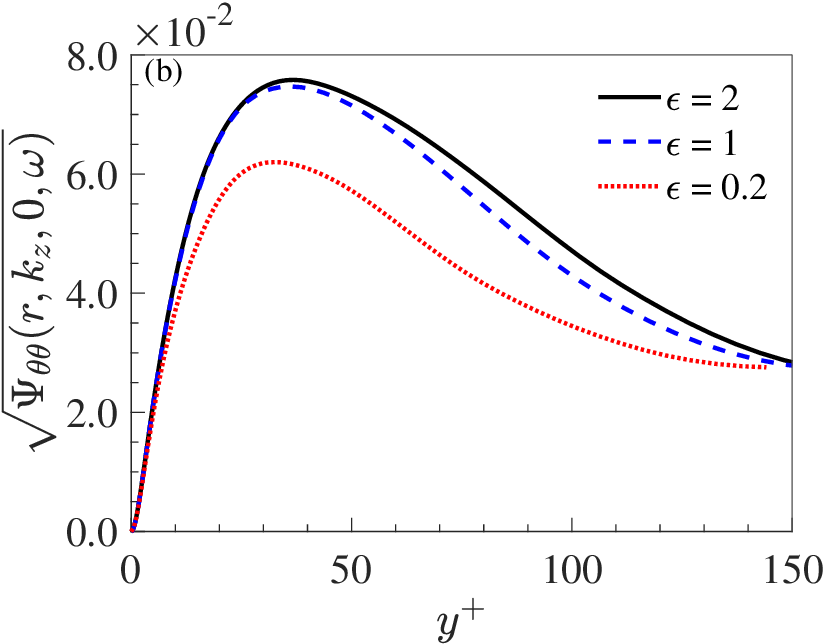}
    \end{subfigure}
    \caption{Wall normal distribution of the Lighthill stress components (a)
    $\tilde{T}_{rr}$ and (b) $\tilde{T}_{\theta\theta}$ over inner cylinder
    surface at $k_z\delta=0$ and $\omega\delta/U_b\approx0.165$.}
    \label{fig:TijInnerKxInd1f0p32}
\end{figure}

Figure~\ref{fig:TijInnerKxInd1f0} shows the amplitude spectra of
$\tilde{T}_{rr}(r, 0, 0, 0)$ and $\tilde{T}_{\theta\theta}(r, 0, 0, 0)$. These
are the steady and homogeneous components of the Lighthill stress tensors and
should be more properly interpreted as propagation effects rather than as a
source in the acoustic analogy framework. Nevertheless, we see that as curvature
increases, the two steady stress components generally decrease in amplitude.
However, they do not differ from each other by orders of magnitude. In terms of
decibels the difference in their amplitudes is limited by 3 dB. 
\begin{figure}
    \centering
    \begin{subfigure}{0.495\textwidth}
    \includegraphics[width=\linewidth]{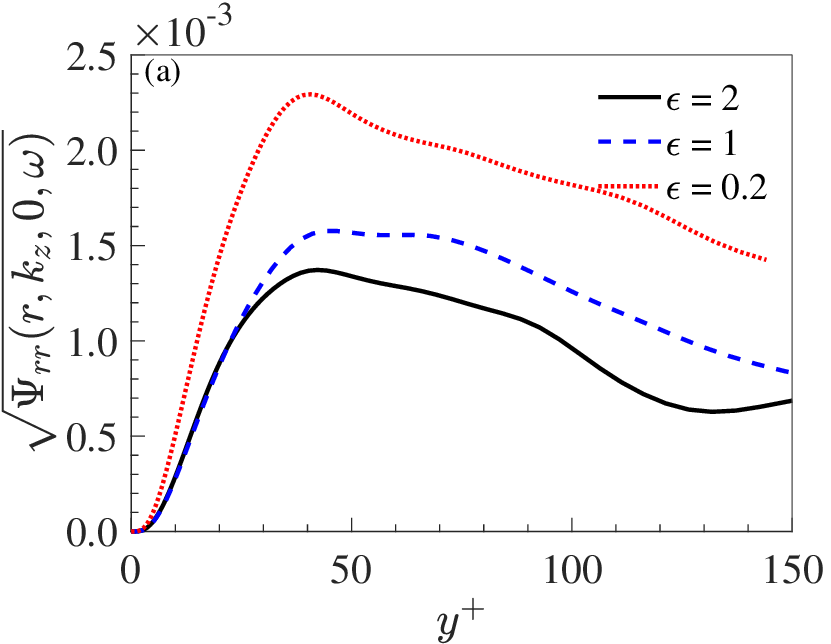}
    \end{subfigure}
    \begin{subfigure}{0.495\textwidth}
    \includegraphics[width=\linewidth]{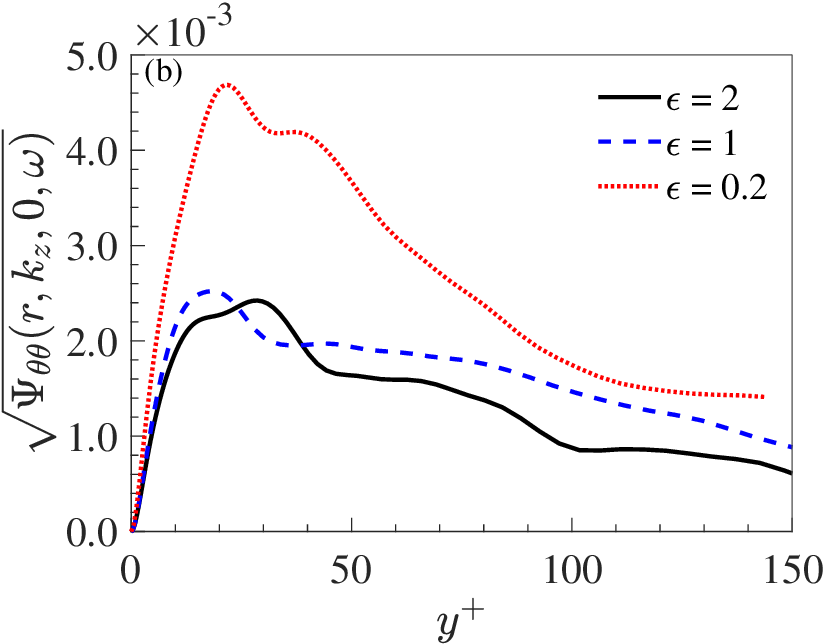}
    \end{subfigure}
    \caption{Wall normal distribution of the Lighthill stress components (a)
    $\tilde{T}_{rr}$ and (b) $\tilde{T}_{\theta\theta}$ over inner cylinder
    surface at $k_z\delta=0$ and $\omega\delta/U_b\approx0.33$.}
    \label{fig:TijInnerKxInd1f0p064}
\end{figure}
\begin{figure}
    \centering
    \begin{subfigure}{0.495\textwidth}
    \includegraphics[width=\linewidth]{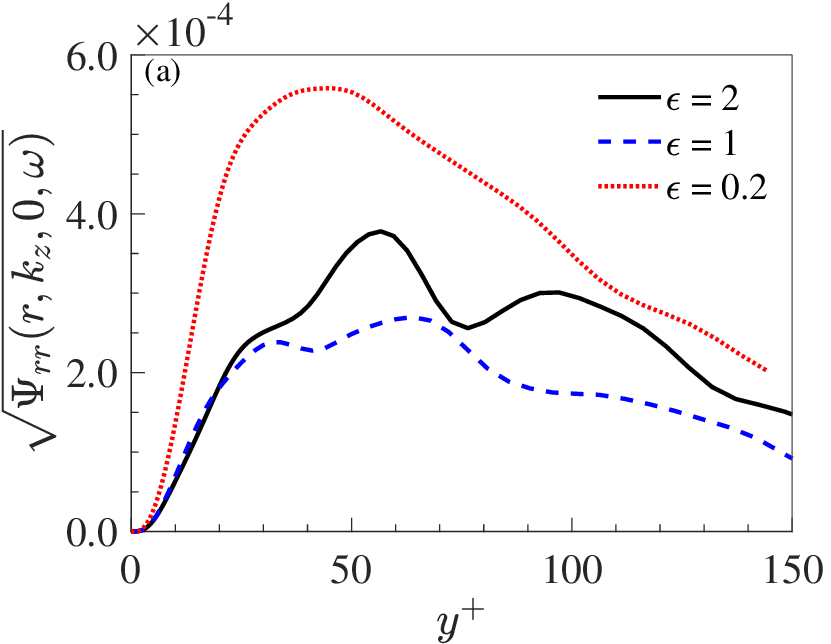}
    \end{subfigure}
    \begin{subfigure}{0.495\textwidth}
    \includegraphics[width=\linewidth]{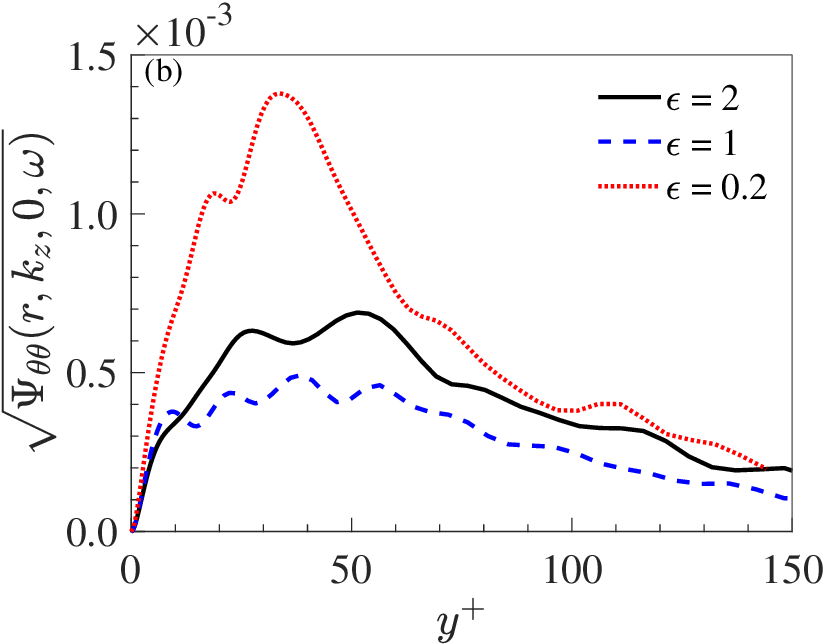}
    \end{subfigure}
    \caption{Wall normal distribution of the Lighthill stress components (a)
    $\tilde{T}_{rr}$ and (b) $\tilde{T}_{\theta\theta}$ over inner cylinder
    surface at $k_z\delta=0$ and $\omega\delta/U_b\approx 1.16$.}
    \label{fig:TijInnerKxInd1f0p2}
\end{figure}

The distribution at $k_z \delta=0$ and $\omega \delta / U_b \approx 0.165$, as
demonstrated in figure~\ref{fig:TijInnerKxInd1f0p32}, shows a similar story.
However, it is worth noting that the amplitude becomes much smaller than that at
$\omega \delta/U_b=0$. Such a trend further continues as frequency increases, as
shown in figures~\ref{fig:TijInnerKxInd1f0p064} and
\ref{fig:TijInnerKxInd1f0p2}, where the non-dimensional frequency $\omega \delta
/ U_b \approx 0.33$ and $1.16$, respectively. This is consistent with the fact
the low-wavenumber WPF decreases quickly as frequency increases. Unlike
figures~\ref{fig:TijInnerKxInd1f0} and \ref{fig:TijInnerKxInd1f0p32},
figures~\ref{fig:TijInnerKxInd1f0p064} and \ref{fig:TijInnerKxInd1f0p2} show
that the amplitudes of $\tilde{T}_{\theta\theta}$ and $\tilde{T}_{rr}$ largely
increase as the curvature increases. However, the difference is not significant.
Distributions of these two stress components at even higher frequencies are
shown in Appendix B. It is worth noting that as frequency increases, the
Lighthill stress components appear to converge increasingly less satisfactorily.


We can now calculate both $\tilde{T}_{rr}$ and $\tilde{T}_{\theta\theta}$ and
predict the wavenumber-frequency spectra of the WPF using
(\ref{equ:0modeCompact}). Note that the model developed in this section is for
the WPF beneath a homogeneous turbulent boundary layer; therefore, it is not
strictly applicable to the present channel flows. Nevertheless, we can still
compare the model prediction with the simulation results, hoping to capture the
important trend instead of seeking an exact agreement. 

Results are shown in figure~\ref{fig:ComparisonOmega}, where the predicted WPF
wavenumber-frequency spectra $\Phi(k_z, m, \omega)$ for the axisymmetric ($m=0$) mode at low frequencies are compared with DNS results. Again, since we are
primarily interested in the low-wavenumber where the WPF has a ``wavenumber
white'' behaviour near $k_z\delta=0$, we show the spectrum $\Phi(0, 0, \omega)$
as a function of $\omega$. The predicted spectra are obtained from
(\ref{equ:0modeCompact}) using $\tilde{T}_{rr}$ and $\tilde{T}_{\theta\theta}$
examined in figures~\ref{fig:TijInnerKxInd1f0} to \ref{fig:TijInnerKxInd1f0p2}.
From figure~\ref{fig:ComparisonOmega}, we can see that spectral augmentation as
$\epsilon$ decreases is well-predicted from the asymptotic equation. As
$\epsilon$ decreases from $2$ to $0.2$, the amplitude of the DNS WPF spectra
increases by almost two orders of magnitude. This trend is well captured in the
prediction. Considering that $\tilde{T}_{r r}$ and $\tilde{T}_{\theta\theta}$ do
not change significantly as $\epsilon$ varies, this augmentation is indeed
primarily contributed by the $\epsilon+h$ term in (\ref{equ:0modeCompact}). 

In addition, outside the acoustic peaks, the predicted spectra follow the DNS
results closely. The agreement is slightly better for $\epsilon=2$ and
$\epsilon=1$, while the spectrum of $\epsilon=0.2$ is slightly over-predicted.
Note that decreasing the upper interval limit of the integral in
(\ref{equ:0modeCompact}) slightly reduces the predicted $\Phi$ amplitude
significantly for $\epsilon=2$ and $1$, but the variation is negligible for
$\epsilon=0.2$. This is a direct demonstration that disturbances closer to the
wall are increasingly important in generating WPF as the inner cylinder becomes
thinner. Considering that the model does differ from the channel flow setup,
such an agreement is sufficient to show the validity of
(\ref{equ:0modeCompact}). 

\begin{figure}
    \centering
    \includegraphics[width=0.52\textwidth]{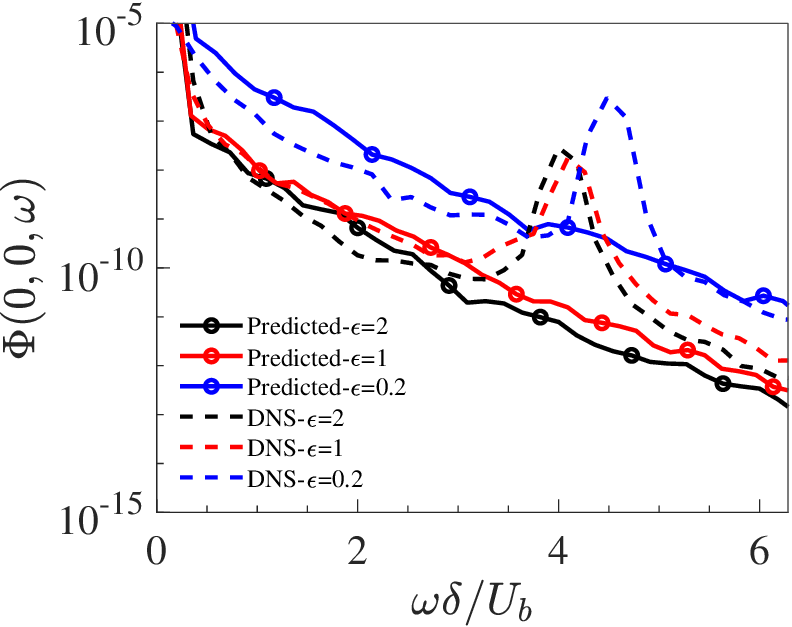}
    \caption{Comparison of the WPF wavenumber frequency spectrum $\Phi(0, 0,
    \omega)$ between the asymptotic prediction and DNS results.}
    \label{fig:ComparisonOmega}
\end{figure}
}

\section{Conclusion and future work}\label{sec:conclusion}
Compressible DNS simulations within concentric annular ducts are performed in
this paper to study the effects of curvature on the WPF characteristics over
cylindrical walls. The inner cylinder radius $R$ ranges from $0.2\delta$,
$\delta$, $2\delta$ to $\infty$. \add{The simulations are validated against
previous results in the literature, and the effects of curvature on the flow
statistics are discussed. It is found that, as the curvature increases, the
three normal  Reynolds stress components, together with $\langle u_r^{\prime
+}u_z^{\prime +}\rangle$, over the inner cylinder wall decrease and peak
slightly closer to the wall. However, the flows over the outer cylinder wall
show an opposite behaviour.} 

\add{Regarding the WPF, results show that, as $R$ decreases, the magnitude of
the one-point PSD of the WPF decreases at intermediate but increases at high
frequencies.} The magnitude of the 1D streamwise wavenumber-frequency spectrum
decreases at high wavenumbers. Moreover, as $R$ reduces to $0.2\delta$, the
spectra start to show acoustic peaks at low wavenumbers, whose strengths become
increasingly strong at higher frequencies. \add{These acoustic peaks represent
acoustic duct modes within the concentric annular channel. An examination of
these modes shows that the plane wave mode (1st-order radial mode) does not
discriminate between the inner and outer walls in terms of pressure amplitude,
however, modes of higher radial orders are increasingly more energetic on the
inner walls.}

The 0th-order 2D wavenumber-frequency spectrum is of great importance in
practical applications. DNS shows that reducing $R$ decreases the
high-wavenumber but increases the low-wavenumber components. In particular, the
low-wavenumber increase appears stronger at high frequencies. Analytical
modelling based on the Lighthill acoustic analogy shows that this arises
primarily from the ``geometric'' effects connected with the Green's function.
\add{Asymptotic prediction of this model agrees well with the DNS results at low
frequencies.} The 0th-order low-wavenumber WPF is generated mainly by azimuthal
and radial disturbances. Moreover, disturbances close to the wall play an
increasingly important role in generating WPF as the curvature increases. This
is important as it suggests a possibility of low-wavenumber WPF control using
wall treatments. At $R=0.2\delta$, 2D wavenumber-frequency spectra of higher
azimuthal orders are virtually identical to the 0th-order except nearby acoustic
ridges.

The present work mainly focuses on the effects of curvature on the WPF
characteristics. The effects of Mach number are examined in \citet{Liu2024} for
rectangular channels. It is, however, worth studying how these effects change in
the annular channel. The difference in WPF between channel flows and free
boundary layers also warrants investigations. These form part of our future
work.

\backsection[Acknowledgements]{The authors wish to thank Prof. Richard D. Sandberg for the license of using the code HiPSTAR, and Prof. Meng Wang for inspiring discussions on compressible DNS simulations.}

\backsection[Funding]{The authors wish to gratefully acknowledge the National
Natural Science Foundation of China (NSFC) under grant numbers 12472263,
U25700222, 92152202, 12432010, 12588201 and 12572247.}
\backsection[Declaration of interests]{The authors report no conflict of
interest.}

\appendix
\add{
\section{Distribution of other Reynolds stress components}
\begin{figure}
    \centering
    \begin{subfigure}{0.495\textwidth}
    \includegraphics[width=\linewidth]{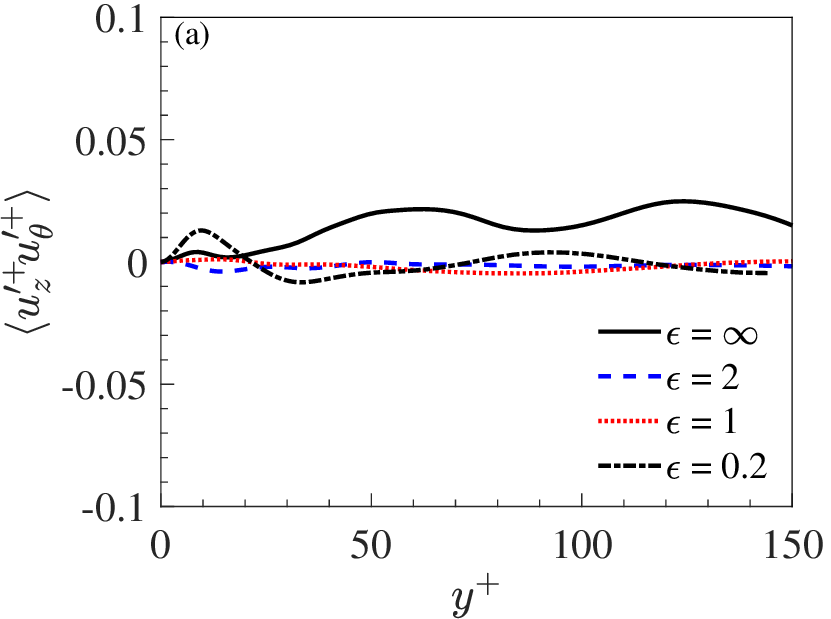}
    \end{subfigure}
    \begin{subfigure}{0.495\textwidth}
    \includegraphics[width=\linewidth]{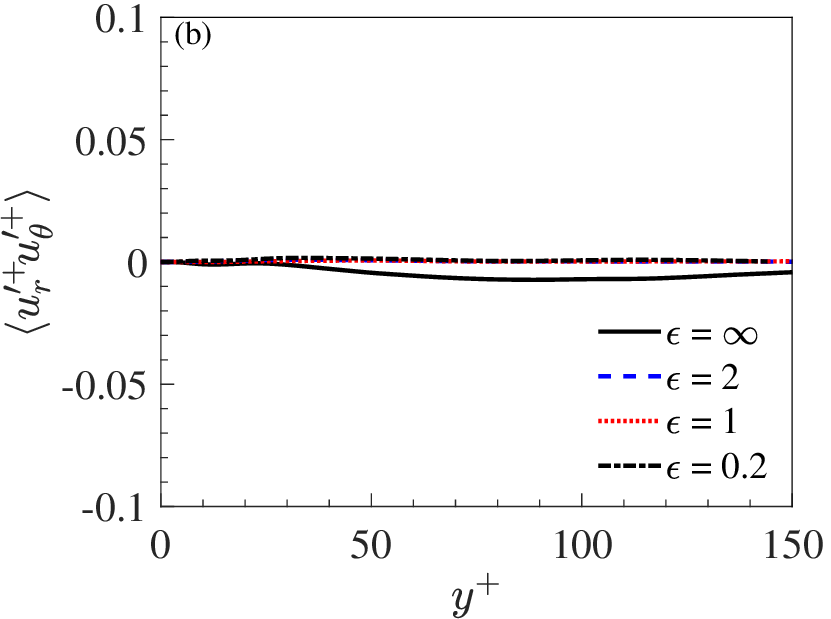}
    \end{subfigure}
    \begin{subfigure}{0.495\textwidth}
    \includegraphics[width=\linewidth]{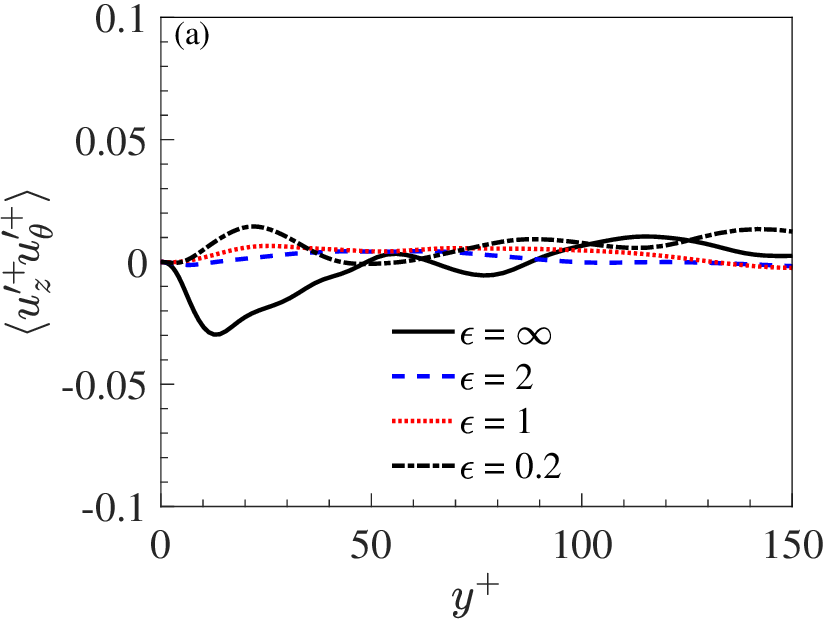}
    \end{subfigure}
    \begin{subfigure}{0.495\textwidth}
    \includegraphics[width=\linewidth]{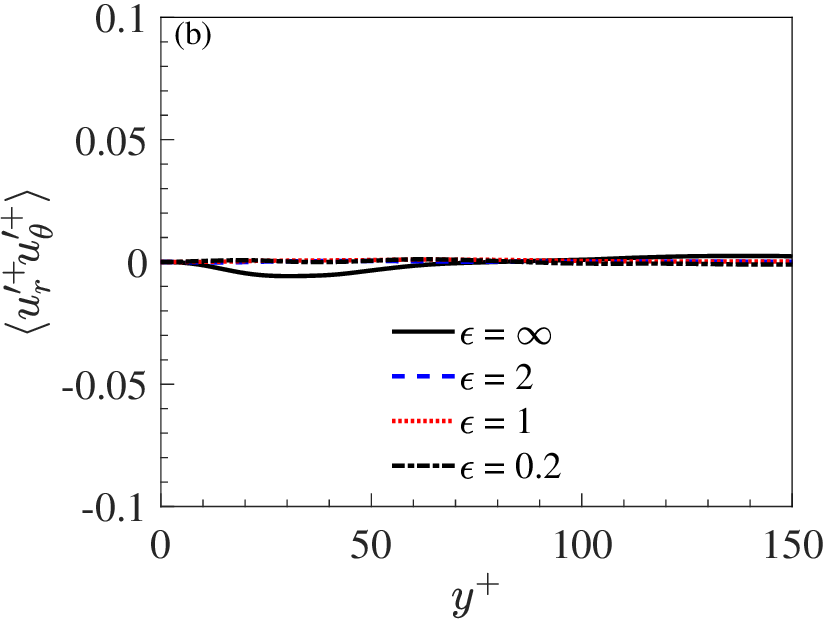}
    \end{subfigure}
    \caption{Wall normal distribution of the Reynolds stresses of (a) $\langle
    u^{\prime+}_zu^{\prime+}_\theta\rangle$ and (b) $\langle
    u^{\prime+}_ru^{\prime+}_\theta\rangle$ over inner cylinder wall and (c)
    $\langle u^{\prime+}_z u^{\prime+}_\theta\rangle$ and (d) $\langle
    u^{\prime+}_r u^{\prime+}_\theta\rangle$ over outer cylinder wall.}
    \label{fig:ReynoldsLastTwo}
\end{figure}
The wall-normal distribution of the other two Reynolds stress components, i.e.
$\langle u_z^{\prime +} u_\theta^{\prime +}\rangle$ and $\langle u_r^{\prime +}
u_\theta^{\prime +}\rangle$ over the inner and outer cylinder walls are shown in
figure~\ref{fig:ReynoldsLastTwo}. Among them,
figures~\ref{fig:ReynoldsLastTwo}(a) and (b) show the two components over the
inner wall. First of all, as expected, both $\langle u_r^{\prime +}
u_\theta^{\prime +}\rangle$ and $\langle u_z^{\prime +} u_\theta^{\prime
+}\rangle$ are two orders of magnitude smaller than other Reynolds stress
components, while the former appears larger than the latter. In addition, the
plane channel flow appears to have the largest value. The wall-normal
distributions over the outer cylinder wall, as shown in
figures~\ref{fig:ReynoldsLastTwo}(c) and (d), are generally similar. It is also
worth mentioning that neither component contributes to the 0th-order WPF
wavenumber-frequency spectra, as shown in section~\ref{sec:discussions}.

\section{Distribution of other Lighthill stress components}
\begin{figure}
    \centering
    \begin{subfigure}{0.495\textwidth}
    \includegraphics[width=\linewidth]{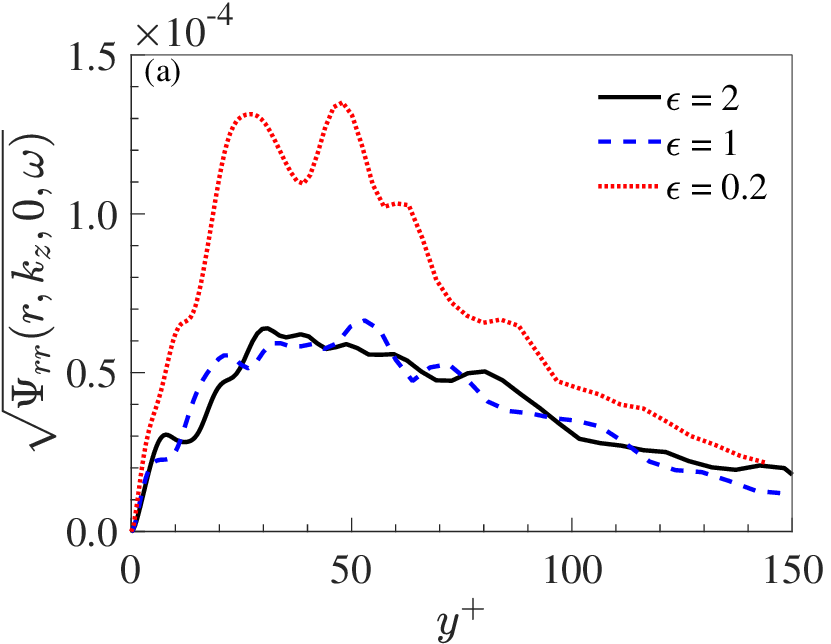}
    \end{subfigure}
    \begin{subfigure}{0.495\textwidth}
    \includegraphics[width=\linewidth]{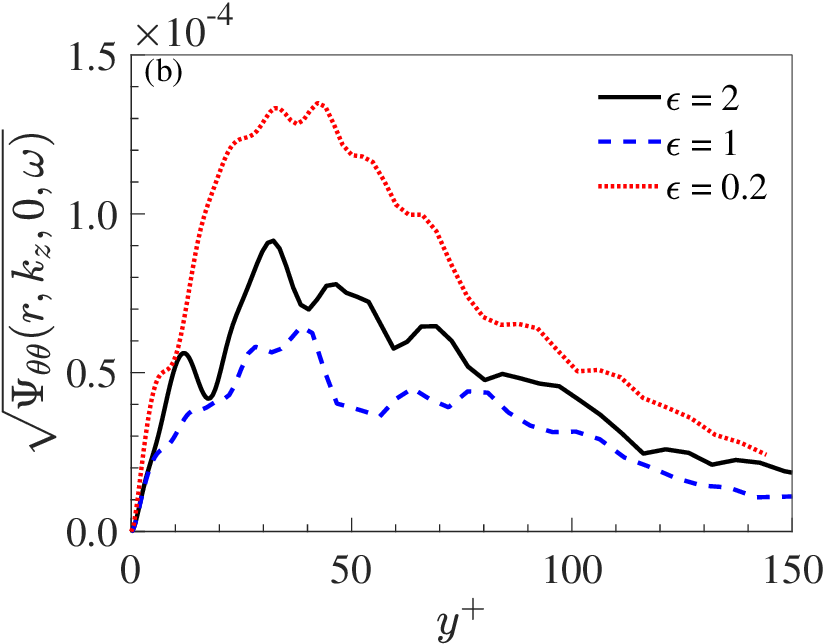}
    \end{subfigure}
    \caption{Wall normal distribution of the Lighthill stress components (a)
    $\tilde{T}_{rr}$ and (b) $\tilde{T}_{\theta\theta}$ over inner cylinder
    surface at $k_z\delta=0$ and $\omega \delta / U_b=\pi$.}
    \label{fig:TijInnerKxInd1f0p5}
\end{figure}
\begin{figure}
    \centering
    \begin{subfigure}{0.495\textwidth}
    \includegraphics[width=\linewidth]{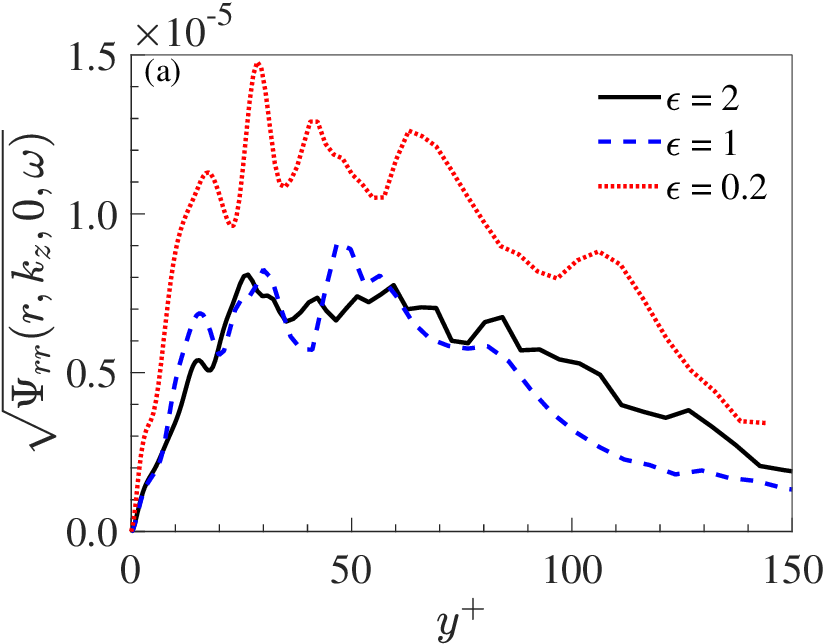}
    \end{subfigure}
    \begin{subfigure}{0.495\textwidth}
    \includegraphics[width=\linewidth]{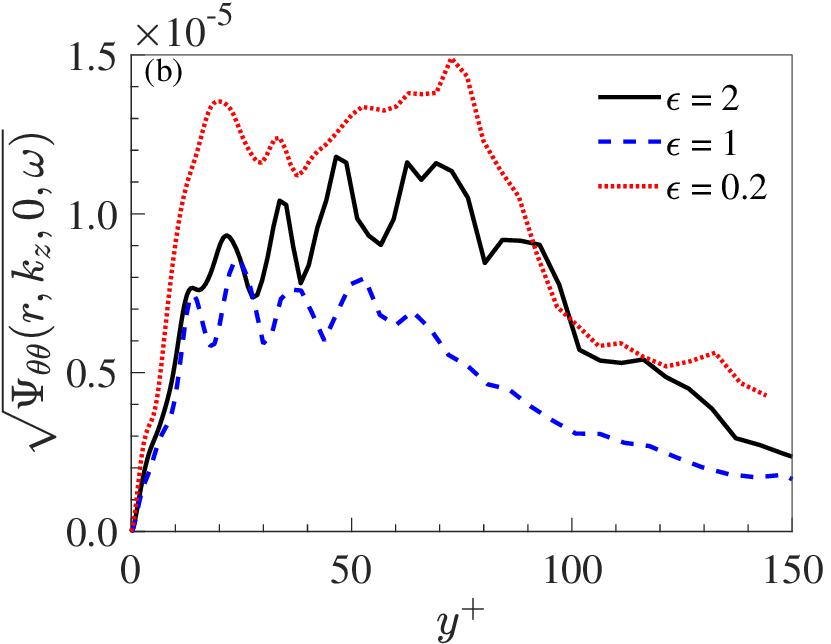}
    \end{subfigure}
    \caption{Wall normal distribution of the Lighthill stress components (a)
    $\tilde{T}_{rr}$ and (b) $\tilde{T}_{\theta\theta}$ over inner cylinder
    surface at $k_z\delta=0$ and $\omega \delta / U_b=2\pi$.}
    \label{fig:TijInnerKxInd1f1}
\end{figure}
To further examine the assumption that $\tilde{T}_{\theta\theta}$ and
$\tilde{T}_{rr}$ do not change significantly as curvature varies, we also show
the distribution of them at $\omega\delta/U_b=\pi$ and $2\pi$ in
figure~\ref{fig:TijInnerKxInd1f0p5} and \ref{fig:TijInnerKxInd1f1},
respectively. These correspond to $k\delta$ values of $0.2$ and $0.4$,
respectively. The case of $\omega \delta / U_b=\pi$ shown in
figure~\ref{fig:TijInnerKxInd1f0p5} is similar to that in
figure~\ref{fig:TijInnerKxInd1f0p2}, where the amplitudes of $\tilde{T}_{rr}$
and $\tilde{T}_{\theta\theta}$ grows larger when $\epsilon=0.2$. However, the
difference is not significant. Examining these distributions at $\omega \delta /
U_b=2\pi$ in figure~\ref{fig:TijInnerKxInd1f1}, we note first that the
amplitudes of both components have decayed significantly to $10^{-5}$. In
addition, one can see that as curvature increases, both
$\tilde{T}_{\theta\theta}$ and $\tilde{T}_{rr}$ increase only slightly. The
difference between them is small. From figures~\ref{fig:TijInnerKxInd1f0} to
\ref{fig:TijInnerKxInd1f1}, it is clear that these Lighthill stress components
indeed do not differ significantly at the low frequencies. }

\bibliography{cleanRef}

\begin{thebibliography}{29}
\expandafter\ifx\csname natexlab\endcsname\relax\def\natexlab#1{#1}\fi
\def\au#1{#1} \def\ed#1{#1} \def\yr#1{#1}\def\at#1{#1}\def\jt#1{\textit{#1}} \def\bt#1{#1}\def\bvol#1{\textbf{#1}} \def\vol#1{#1} \def\pg#1{#1} \def\publ#1{#1}\def\arxiv#1{#1}\def\org#1{#1}\def\st#1{\textit{#1}}

\bibitem[Bagheri \& Wang(2020)]{BAGHERI2020}
{\sc \au{Bagheri, E.} \& \au{Wang, B.-C.}} \yr{2020}  \at{Effects of radius ratio on turbulent concentric annular pipe flow and structures}.  \jt{International Journal of Heat and Fluid Flow}  \bvol{86},  \pg{108725}.

\bibitem[Blake(1970)]{blake_turbulent_1970}
{\sc \au{Blake, W.~K.}} \yr{1970}  \at{Turbulent boundary-layer wall-pressure fluctuations on smooth and rough walls}.  \jt{Journal of Fluid Mechanics}  \bvol{44}~(4),  \pg{637--660}.

\bibitem[Bogey \& Bailly(2004)]{bogey2004family}
{\sc \au{Bogey, Christophe} \& \au{Bailly, Christophe}} \yr{2004}  \at{A family of low dispersive and low dissipative explicit schemes for flow and noise computations}.  \jt{Journal of Computational physics}  \bvol{194}~(1),  \pg{194--214}.

\bibitem[Bull(1967)]{Bull1967}
{\sc \au{Bull, M.~K.}} \yr{1967}  \at{Wall pressure fluctuations associated with subsonic turbulent boundary layer flow}.  \jt{Journal of Fluid Mechanics}  \bvol{28},  \pg{719--7544}.

\bibitem[Bull(1996)]{Bull1996}
{\sc \au{Bull, M.~K}} \yr{1996}  \at{Wall-pressure fluctuations beneath turbulent boundary layers: some reflections on forty years of research}.  \jt{Journal of Sound and Vibration}  \bvol{190},  \pg{299--315}.

\bibitem[Cohen \& Gloerfelt(2018)]{Cohen2018}
{\sc \au{Cohen, E.} \& \au{Gloerfelt, X.}} \yr{2018}  \at{Influence of pressure gradients on wall pressure beneath a turbulent boundary layer}.  \jt{Journal of Fluid Mechanics}  \bvol{838},  \pg{715--758}.

\bibitem[Corcos(1963)]{Corcos1963}
{\sc \au{Corcos, G.~M.}} \yr{1963}  \at{Resolution of pressure in turbulence}.  \jt{Journal of Acoustical Society of America}  \bvol{35},  \pg{192--199}.

\bibitem[Crighton {\em et~al.\/}(1992)Crighton, Dowling, Ffowcs‐Williams, Heckl, Leppington \& Bartram]{crighton_modern_1992}
{\sc \au{Crighton, David~George}, \au{Dowling, Ann~P}, \au{Ffowcs‐Williams, J.~E.}, \au{Heckl, Manfred}, \au{Leppington, FG} \& \au{Bartram, James~F}} \yr{1992}  \at{Modern methods in analytical acoustics lecture notes} Publisher: Acoustical Society of America.

\bibitem[Deuse \& Sandberg(2020)]{deuse2020different}
{\sc \au{Deuse, Mathieu} \& \au{Sandberg, Richard~D.}} \yr{2020}  \at{Different noise generation mechanisms of a controlled diffusion aerofoil and their dependence on mach number}.  \jt{Journal of Sound and Vibration}  \bvol{476},  \pg{115317}.

\bibitem[Dhanak(1988)]{Dhanak1988}
{\sc \au{Dhanak, M.~R.}} \yr{1988}  \at{Turbulent boundary layer on circular cylinder: the low-wavenumber surface pressure spectrum due to a low-mach number flow}.  \jt{Journal of Fluid Mechanics}  \bvol{191},  \pg{443--464}.

\bibitem[Dowling(1998)]{dowling_underwater_1998}
{\sc \au{Dowling, A.P.}} \yr{1998}  \at{Underwater {flow} {noise}}.  \jt{Theoretical and Computational Fluid Dynamics}  \bvol{10}~(1),  \pg{135--153}.

\bibitem[Ffowcs-Williams(1965)]{williams_surface-pressure_1965}
{\sc \au{Ffowcs-Williams, J.~E.}} \yr{1965}  \at{Surface-pressure fluctuations induced by boundary-layer flow at finite {Mach} number}.  \jt{Journal of Fluid Mechanics}  \bvol{22}~(3),  \pg{507--519}, publisher: Cambridge University Press.

\bibitem[Gloerfelt \& Berland(2013)]{Gloerfelt2013}
{\sc \au{Gloerfelt, X.} \& \au{Berland, J.}} \yr{2013}  \at{Turbulent boundary-layer noise: direct radiation at mach number 0.5}.  \jt{Journal of Fluid Mechanics}  \bvol{723},  \pg{318--351}.

\bibitem[Harrison(1958)]{Harrison1958}
{\sc \au{Harrison, M.}} \yr{1958}  \bt{Pressure fluctuations on the wall adjacent to a turbulent boundary layer}. Report No. 1260.  \org{U.S. Navy David Taylor Model Basin}.

\bibitem[Howe(1987)]{Howe1987}
{\sc \au{Howe, M.~S.}} \yr{1987}  \at{On the structure of the turbulent boundary-layer wall pressure spectrum in the vicinity of the acoustic wavenumber}.  \jt{Proceedings of the Royal Society, London A}  \bvol{412},  \pg{389--401}.

\bibitem[Kennedy {\em et~al.\/}(2000)Kennedy, Carpenter \& Lewis]{kennedy2000low}
{\sc \au{Kennedy, C.~A.}, \au{Carpenter, M.~H.} \& \au{Lewis, R.~M.}} \yr{2000}  \at{{Low-storage, explicit Runge--Kutta schemes for the compressible Navier--Stokes equations}}.  \jt{Appl. Numer. Math.}  \bvol{35}~(3),  \pg{177--219}.

\bibitem[Kennedy \& Gruber(2008)]{kennedy2008reduced}
{\sc \au{Kennedy, Christopher~A.} \& \au{Gruber, Andrea}} \yr{2008}  \at{Reduced aliasing formulations of the convective terms within the navier--stokes equations for a compressible fluid}.  \jt{Journal of Computational Physics}  \bvol{227}~(3),  \pg{1676--1700}.

\bibitem[Kraichnan(1956)]{kraichnan_pressure_1956}
{\sc \au{Kraichnan, Robert~H}} \yr{1956}  \at{Pressure fluctuations in turbulent flow over a flat plate}.  \jt{The Journal of the Acoustical Society of America}  \bvol{28}~(3),  \pg{378--390}, publisher: Acoustical Society of America.

\bibitem[Liu {\em et~al.\/}(2024)Liu, Wang \& Wang]{Liu2024}
{\sc \au{Liu, Y.}, \au{Wang, K.} \& \au{Wang, M.}} \yr{2024}  \at{Subconvective wall-pressure fluctuations in low-mach-number turbulent channel flow}.  \jt{Journal of Fluid Mechanics}  \bvol{984}~(R2),  \pg{1--13}.

\bibitem[Neves \& Moin(1994{\natexlab{{\em a\/}}})]{Neves1994a}
{\sc \au{Neves, J.~C.} \& \au{Moin, P.}} \yr{1994{\natexlab{{\em a\/}}}}  \at{Effects of convex transverse curvature on wall-bounded turbulence. part 1. velocity and vorticity}.  \jt{Journal of Fluid Mechanics}  \bvol{272},  \pg{349--381}.

\bibitem[Neves \& Moin(1994{\natexlab{{\em b\/}}})]{Neves1994b}
{\sc \au{Neves, J.~C.} \& \au{Moin, P.}} \yr{1994{\natexlab{{\em b\/}}}}  \at{Effects of convex transverse curvature on wall-bounded turbulence. part 2. the pressure fluctautions}.  \jt{Journal of Fluid Mechanics}  \bvol{272},  \pg{383--406}.

\bibitem[Pope(2000)]{Pope2000}
{\sc \au{Pope, S.~B.}} \yr{2000} {\em Turbulent Flows\/}.  \publ{Cambridge University Press}.

\bibitem[Sandberg {\em et~al.\/}(2015)Sandberg, Michelassi, Pichler, Chen \& Johnstone]{sandberg2015compressible}
{\sc \au{Sandberg, R.~D.}, \au{Michelassi, V.}, \au{Pichler, R.}, \au{Chen, L.} \& \au{Johnstone, R.}} \yr{2015}  \at{Compressible direct numerical simulation of low-pressure turbines---{Part I}: Methodology}.  \jt{J. Turbomach.}  \bvol{137}~(5),  \pg{051011}.

\bibitem[White(1991)]{white1991viscous}
{\sc \au{White, F.~M.}} \yr{1991} {\em Viscous fluid flow\/}.  \publ{McGraw-Hill, New York}.

\bibitem[Willmarth(1956)]{Willmarth1956}
{\sc \au{Willmarth, W.~W.}} \yr{1956}  \at{Wall pressure fluctuations in a turbulent boundary layer}.  \jt{Journal of Acoustical Society of America}  \bvol{28},  \pg{1048--1053}.

\bibitem[Willmarth(1958)]{Willmarth1958}
{\sc \au{Willmarth, W.~W.}} \yr{1958}  \at{Space-time correlations of the fluctuating wall pressure in a turbulent boundary layer}.  \jt{Journal of Aerospace Science}  \bvol{25},  \pg{335--336}.

\bibitem[Willmarth(1975)]{Willmarth1975}
{\sc \au{Willmarth, W.~W.}} \yr{1975}  \at{Pressure fluctuations beneath turbulent boundary layers}.  \jt{Annual Review of Fluid Mechanics}  \bvol{7},  \pg{13--36}.

\bibitem[Willmarth {\em et~al.\/}(1976)Willmarth, Winkel, Sharma \& Bogar]{Willmarth1976}
{\sc \au{Willmarth, W.~W.}, \au{Winkel, R.~E.}, \au{Sharma, L.~K.} \& \au{Bogar, T.~J.}} \yr{1976}  \at{Axially symmetric turbulent boundary layers on cylinders: Mean velocity profiles and wall pressure fluctuations}.  \jt{Journal of Fluid Mechanics}  \bvol{76},  \pg{35--64}.

\bibitem[Willmarth \& Yang(1970)]{Willmarth1970}
{\sc \au{Willmarth, W.~W.} \& \au{Yang, C.~S.}} \yr{1970}  \at{Wall pressure fluctuations beneath turbulent boundary layer on a flat plate and a cylinder}.  \jt{Journal of Fluid Mechanics}  \bvol{41},  \pg{47--80}.

\end{thebibliography}
\bibliographystyle{jfm}

\end{document}